\documentstyle[preprint,eqsecnum,aps]{revtex}
\newif\iftightenlines\tightenlinesfalse
\tightenlines\tightenlinestrue
\begin{document}
%
%%%%%%%%%%%%%%%%%%%%%%%%%%%%%%%%%%%%%%%%%%%%%%%%%%%%%%%%%%%%%%%%%
%
%%%%%%%%%%%%%%%%%%%% TITLE PAGE %%%%%%%%%%%%%%%%%%%%%%%%%%%%%%%%%
%
\draft
\preprint{
   \vbox{\baselineskip=14pt%
   \rightline{ITP-SB-01-98}\break
   \rightline{INLO-PUB-01/98} }
}

\title{
TWO-LOOP OPERATOR MATRIX ELEMENTS CALCULATED UP TO FINITE TERMS
}

\author{
Y. Matiounine$^1$, J. Smith$^1$ and W.L. van Neerven$^2$
}

\address{
$^1$Institute for Theoretical Physics,
State University of New York at Stony Brook,
Stony Brook, NY 11794-3840, USA
}
\address{
$^2$Instituut-Lorentz,
University of Leiden,
PO Box 9506, 2300 RA Leiden, The Netherlands
}

%\footnote{ Partially supported under the contract NSF PHY-9722101.}  }

\date{January 1998}
\maketitle
\begin{abstract}
We present the two-loop corrected operator matrix elements 
calculated in $N$-dimensional regularization up to the finite terms 
which survive in the limit $\varepsilon = N - 4 \rightarrow 0 $.
The anomalous dimensions of the local operators have been 
previously extracted from the pole terms and 
determine the scale evolution of the deep inelastic
structure functions measured in unpolarized lepton hadron scattering. 
The finite $\varepsilon$-independent 
terms in the two-loop expressions are needed to renormalize the 
local operators up to third order in the strong coupling constant $\alpha_s$. 
Further the unrenormalized
expressions for the two-loop corrected operator matrix elements can be inserted
into specific one loop graphs to obtain a part of the third order 
contributions to these matrix elements. 
This work is a first step in obtaining the anomalous dimensions
up to third order so that a complete next-to-next-to-leading order 
(NNLO) analysis
can be carried out for deep inelastic electroproduction.
\end{abstract}
\medskip
\pacs{PACS numbers : 11.15.Bt, 12.38.-t, 12.38.Bx}
%%%%%%%%%%%%%%%%%%%%%%%%%%%%%%%%%%%%%%%%%%%%%%%%%%%%%%%
%% MAIN TEXT %%
%\newcommand{\mysection}{\setcounter{equation}{0}\section}
%------------------This is Section 1---------------------------------
\section{Introduction}
%----------------------------------------------------------

One of the most important successes of the theory of perturbative Quantum 
Chromodynamics (QCD) is the prediction of the scale evolution of the
structure functions ${\rm F}_i(x,Q^2)$ measured in deep inelastic lepton-hadron
scattering (for reviews see \cite{pol}).
From these structure functions one can infer the parton densities, which
serve as input for many other deep-inelastic (hard) processes, and lead to
a wealth of predictions for cross sections (for a review see \cite{ster}).
Many of these predictions have been 
tested by now and are in very good agreement with the experimental data.
The scale evolution of the structure functions is determined by the
anomalous dimensions of local composite operators $O_i^n(x)$, $i=q,g$,
where $n$ denotes the spin, which show up
in the expansion of the product of two local electroweak 
currents ${\rm J}_{\mu}(x)$ and 
${\rm J}_{\nu}(y)$ around the lightcone $(x-y)^2 = 0$. The structure
functions are proportional to the Fourier transform into momentum space
of the product of these two currents sandwiched
between hadronic states. In this way the structure functions can be written
as a product of operator matrix elements (OME's) and coefficient 
functions. The former describe the long 
distance (low momenta) properties of QCD whereas the latter
account for the short distance (large momenta) behaviour of the specific 
quantities, like e.g. the structure functions, under consideration. Both 
the anomalous dimensions and the coefficient functions are calculable order
by order in perturbation theory so that they can be expressed in a series 
expansion in the strong coupling constant $\alpha_s$. The finite
anomalous dimensions are responsible for the scale dependence of the structure
functions and the parton densities. On the other hand the
OME's themselves are of a non-perturbative nature, and,
apart from some attempts using lattice gauge field theory, cannot yet be 
determined from first principles. Hence they have to be
determined from models and fits to experimental data. 
In the past one has put a lot of 
effort into calculating the lowest order coefficients in the perturbation
series for the anomalous dimensions and the coefficient functions. In 
\cite{gepo} and \cite{grwi} the order $\alpha_s$ 
contributions to the anomalous dimensions 
were obtained by evaluating the one-loop 
OME's which follow from inserting the local operators $O_i^n(x)$
between quark and gluon states (for an alternative method see \cite{alpa}). 
The second order contributions, which involved
the evaluation of two-loop OME's, have been computed by various groups 
\cite{frs}-\cite{flko}. 
As far as the coefficient functions are concerned they have been 
evaluated for many processes up to next-to-leading order (NLO) 
(for a recent review see \cite{ster}). Combining them with the above
NLO anomalous dimensions one can make a complete NLO analysis of 
many quantities like cross sections for Drell-Yan production
or deep inelastic structure functions. 

The advent of HERA opened up a new era of deep inelastic lepton-hadron
experiments with much higher statistics than was previously available.
Furthermore the values of $x$ and $Q^2$, on which the structure functions 
depend, could be extended beyond those accessible by 
earlier fixed target experiments.
This enables QCD to be tested with 
even higher degree of precision so that effects beyond NLO can be
studied. In particular both the small and the large $x$-regions have 
attracted much attention
in the literature (for a review see e.g. \cite{hera}). Hence it will be 
necessary to extend the existing expressions for the anomalous dimensions
and the coefficient functions beyond NLO. The first step to obtain the
next-to-next-to-leading order (NNLO) corrections has been made in \cite{zine}
where the order $\alpha_s^2$ corrections to 
the coefficient functions for deep inelastic lepton-hadron scattering
have been computed. The
same program has also been completed for the Drell-Yan process in \cite{hama}. 
For a consistent NNLO analysis we also need the third order contributions
to the anomalous dimensions. Unfortunately the latter are not known yet 
except for those corresponding to the operators $O_i^n(x)$ with spin
$n=2,4,6,8$ and 10. They have been computed in \cite{lnrv} using completely
different methods than those used in \cite{frs}-\cite{flko}. Using the above
second order coefficient functions and the third order anomalous dimensions 
it is possible to make a NNLO analysis of the structure functions $F_2(x,Q^2)$
and $F_3(x,Q^2)$ as long as one limits oneself to the large $x$
($x > 0.01$) and small $Q^2$ (see \cite{kkps}). If one wants to study the 
behaviour of the structure functions at small $x$ or at large $Q^2$ knowledge
of the anomalous dimensions for any spin is indispensable. Since the full
calculation of the latter quantities is a tremendous enterprise it has to
be carried out step by step. The first step is to compute the two-loop
OME's up to finite terms which is an extension of the work 
done in \cite{frs}-\cite{flko}. These finite terms 
are needed to carry out the renormalization
of the three-loop graphs since they determine the single pole terms in 
$N$-dimensional regularization from which one has to extract the third order
contributions to the anomalous dimensions. Moreover by inserting the 
unrenormalized two-loop corrected OME's in one-loop graphs one gets parts
of the expressions for the unrenormalized three-loop graphs. This method
has been used previously to get parts of the two-loop expressions 
by inserting the one-loop corrected OME's into one-loop graphs.

In section 2
we will give an outline of the calculation which will be carried out in 
the Feynman gauge. Although this gauge entails some complications due
to the mixing of physical and unphysical operators it is the only one
in which it is feasible to perform a calculation of Feynman graphs 
for the OME's beyond two-loop order.
All other gauges, like the axial gauge used in \cite{fupe}, lead
to even more complications, which we want to avoid. 
The long expressions obtained
for the full OME's are presented in Appendix A while Appendix B
contains results for non-physical OME's, which are needed to carry
out the renormalization programme.

%\newcommand{\Ds}{\Delta \hspace{-0.52em}/\hspace{0.1em}}
%------------------This is Section 2---------------------------------
\section{The calculation of the two-loop operator matrix elements}
%----------------------------------------------------------
In this section we will give an outline of the calculation of the OME's 
up to two-loop order.
The operators, which appear in unpolarized lepton-hadron scattering, can be
divided into singlet and non-singlet parts 
with respect to the flavour group.
In leading twist (namely twist two) 
the non-singlet quark operator of spin $n$ is given by
\begin{eqnarray}
\label{eqn:2.1}
O_{q,k}^{\mu_1,\mu_2 \cdots \mu_n}(x)=\frac{1}{2} i^{n-1} {\cal S} \Big [ \bar 
\psi(x) \gamma^{\mu_1} D^{\mu_2}
\cdots D^{\mu_n} \frac{\lambda_k}{2} \psi (x)  + \mbox{trace terms}\Big ]\,.
\end{eqnarray}
In the singlet case there are two operators. The quark operator is 
represented by
\begin{eqnarray}
\label{eqn:2.2}
O_q^{\mu_1,\mu_2 \cdots \mu_n}(x)=\frac{1}{2} i^{n-1} {\cal S} \Big [ \bar 
\psi(x) \gamma^{\mu_1} D^{\mu_2}
\cdots D^{\mu_n} \psi (x) + \mbox{trace terms}\Big ]\,,
\end{eqnarray}
and the gluon operator is given by
\begin{eqnarray}
\label{eqn:2.3}
O_g^{\mu_1,\mu_2 \cdots \mu_n}(x)=\frac{1}{2} i^{n-2} {\cal S} \Big [ 
F_{a \alpha}^{\mu_1}(x) D^{\mu_2}
\cdots D^{\mu_{n-1}} F_a^{\alpha \mu_n}(x) + \mbox{trace terms}\Big]\,.
\end{eqnarray}
In these composite operators $\psi$ and $F_a^{\mu\nu}$ stand for the
quark field and the gluon field tensor respectively. 
The $\lambda_k$ in Eq. (\ref{eqn:2.1}) represent the generators of the flavour
group and the index $a$ in Eq. (\ref{eqn:2.3}) stands for the colour. 
Further the above operators are irreducible tensors 
with respect to the Lorentz group so that they have to be
symmetric and traceless in all their Lorentz indices $\mu_i$. 
From the operators above one can derive the Feynman rules for the operator 
vertices in the standard way (see e.g. \cite{frs}, \cite{hane}, \cite{ham}).
This
derivation is facilitated if the operators are multiplied by the source
\begin{eqnarray}
\label{eqn:2.4}
J_{\mu_1\mu_2 \cdots \mu_n} = 
\Delta_{\mu_1}\Delta_{\mu_2}\cdots \Delta_{\mu_n}\,,
\end{eqnarray}
with $\Delta^2=0$ in order to eliminate the trace terms 
in Eqs. (\ref{eqn:2.1})-(\ref{eqn:2.3}). Hence all operator 
vertices in momentum space are multiplied
by a factor $(\Delta \cdot p)^n$. For the computation of the OME's denoted by
\begin{eqnarray}
\label{eqn:2.5}
A_{ij} = \langle j(p) \mid O_i \mid j(p) \rangle
\end{eqnarray}
with $i,j = q,g$ we choose the Feynman gauge except for the one-loop graphs
for which we take the general covariant gauge. For this choice the gluon 
propagator equals
\begin{eqnarray}
\label{eqn:2.6}
\Delta_{ab}^{\mu\nu}(k)= \frac{i \delta_{ab}}{k^2} \Big ( - g^{\mu\nu} +
(1- \xi ) \frac{k^{\mu}k^{\nu}}{k^2} \Big ) \,.
\end{eqnarray}
The matrix element (\ref{eqn:2.5}) has to be considered as a connected Green 
function with the external legs amputated but with the external self 
energies of the partons $j$ included.
In this paper all quarks and gluons are taken to be massless
and the external momentum $p$ is off-shell ($p^2 < 0$) 
in order to get finite expressions for the OME's.
This choice implies that the OME's are not gauge invariant so that they 
cease to be ordinary S-matrix elements.
Moreover they acquire unphysical parts which can be split into two
classes.
The first class originates from the fact that the equations 
of motion (EOM) do not apply anymore, 
which is the case for both non-singlet and singlet OME's. 
The second class can be traced back to the mixing between
so-called gauge invariant (GI) or physical (PHYS) and non-gauge 
invariant (NGI) operators which
originate from the Yang-Mills (here gluonic) sector (see \cite{brs}, 
\cite{dita}, \cite{klzu}, \cite{jole}).
Therefore this second class of unphysical operators only shows
up in the singlet case. The first class does
not contribute to the operator renormalization constants
in contrast to the second class which affects the aforementioned constants
via the mixing between the GI and NGI operators as we will see
below. The way to deal with this mixing is 
described in \cite{hane}, \cite{ham}, \cite{cosc} and \cite{hasm}.

The calculation of the Feynman graphs corresponding to the physical operators
of Eqs. (\ref{eqn:2.1})-(\ref{eqn:2.3}), which are depicted 
in the figures in \cite{frs}, proceeds in the standard way. 
The figures and definitions for the corresponding unphysical operators
are given in \cite{hane}. The Feynman integrals reveal
ultraviolet divergences which are regularized using the method of 
$N$-dimensional regularization. In this way the above divergences show up
in the form of pole terms of the type $(1/\varepsilon)^k$ with $\varepsilon=
N-4$. In \cite{frs}-\cite{flko} it was sufficient to evaluate the one-loop
graphs up to finite terms and the two-loop graphs up 
to single pole terms in order
to get the second order anomalous dimensions. Here we have to include terms
proportional to $\varepsilon$ in the one-loop expressions and the two-loop
graphs have to be computed up to terms which are finite in the limit
$\varepsilon \rightarrow 0$. The way to compute the two-loop
Feynman integrals up to finite terms is presented in \cite{hane} and in 
Appendix B of \cite{bmsmn}. We used the program FORM \cite{jos} 
to do the necessary algebra. 

To check our results for the Feynman diagrams it is useful to have explicit
expressions for the pole terms in $\varepsilon$. Therefore
we will now present the OME's expressed in renormalization group
coefficients, which are defined in \cite{zine} and \cite{hama}. 
The explicit formulae for the physical and unphysical OME's 
can be found in the Appendices A and B respectively. Further
it is implicitly understood that all quantities in the main text,
in particular the anomalous dimensions $\gamma_{ij}$ ($i,j=q,g$),
are Mellin transforms but to avoid additional indices we do not write
a superscript $n$ to indicate this explicitly. (Another way to interpret
the formulae is that the OME's are given in parton momentum fraction 
($z$) space when the anomalous dimensions are replaced by minus the 
corresponding
Altarelli-Parisi splitting functions and the multiplications in moment
space are replaced by convolutions.) 
We have written the OME's in such a way that all
renormalization group coefficients appearing in the expressions below are
renormalized in the ${\overline {\rm MS}}$-scheme.
Up to order $\alpha_s^2$ the non-singlet and the singlet OME's 
can be decomposed as
\begin{eqnarray}
\label{eqn:2.7}
\hat A_{qq}^r= \Big [ 
{\Delta \hspace{-0.52em}/\hspace{0.1em}}
%\Ds 
\hat A_{qq}^{r,{\rm PHYS}} 
+ 
{p \hspace{-0.52em}/\hspace{0.1em}}
%\ps 
\frac{ \Delta \cdot p}{p^2} 
\hat A_{qq}^{r,{\rm EOM}} \Big ] (\Delta \cdot p)^{n-1} \,,
\end{eqnarray}
where $\hat A_{qq}^{r,{\rm PHYS}}$ and $\hat A_{qq}^{r,{\rm EOM}}$ 
with $r={\rm NS,S}$ stand for the physical and unphysical parts respectively. 
The latter enter due to the breakdown of the equations of motion.
The non-singlet OME can now be expressed into renormalization 
group coefficients as follows
\begin{eqnarray}
\label{eqn:2.8}
\hat A_{qq}^{{\rm NS, PHYS}} &=& 1 + \hat a_s S_{\varepsilon} 
\big(\frac{-p^2}{\mu^2}\big)^{\varepsilon/2}
\Big [ \frac{1}{\varepsilon} \gamma_{qq}^{{\rm NS},(0)} + a_{qq}^{{\rm NS},(1)}
+ \varepsilon a_{qq}^{{\rm NS},\varepsilon,(1)} \Big ]
\nonumber\\[2ex]
&& +{\hat a}_s^2 S_{\varepsilon}^2 \big(\frac{-p^2}{\mu^2}\big)^{\varepsilon}
 \Big[ \frac{1}{\varepsilon^2} \Big 
\{ \frac{1}{2} (\gamma_{qq}^{{\rm NS},(0)})^2
 - \beta_0 \gamma_{qq}^{{\rm NS},(0)} \Big \}
\nonumber\\[2ex]
&& + \frac{1}{\varepsilon}\Big \{ \frac{1}{2} \gamma_{qq}^{{\rm NS},(1)}
- 2 \beta_0 a_{qq}^{{\rm NS},(1)} +
\gamma_{qq}^{{\rm NS},(0)} a_{qq}^{{\rm NS},(1)}
\nonumber\\[2ex]
&& - \hat{\xi}  \frac{d~a_{qq}^{{\rm NS},(1)}}
{d \hat{\xi} } z_{\xi} \Big \}
+ a_{qq}^{{\rm NS},(2)} - 2 \beta_0 a_{qq}^{{\rm NS},\varepsilon,(1)}
\nonumber\\[2ex]
&& + \gamma_{qq}^{{\rm NS},(0)} a_{qq}^{{\rm NS},\varepsilon,(1)}
- \hat{\xi}  \frac{d~a_{qq}^{{\rm NS},\varepsilon,(1)}} {d \hat{\xi} } 
z_{\xi}
 \Big ]_{\hat{\xi}=1} \,,
\end{eqnarray}
where $S_{\varepsilon}$ denotes the spherical factor given by
\begin{eqnarray}
\label{eqn:2.9}
S_{\varepsilon} = {\rm exp}\Big [ \frac{\varepsilon}{2} 
\Big ( \gamma_E - \ln 4\pi 
\Big ) \Big ]\,,
\end{eqnarray}
which originates from $N$-dimensional regularization.
In the expressions above all quantities 
that are unrenormalized with respect to operator, coupling 
constant $\alpha_s$ and gauge constant $\xi$ renormalization are
indicated by a hat. The finite terms are written in such a way 
that after all renormalizations the non logarithmic terms (with respect to  
$\ln(-p^2/\mu^2)$) become equal 
to $a_{qq}^{\rm NS,(k)}$ ($k=1,2$). Further we 
have introduced a shorthand notation for the strong coupling constant
so that
\begin{eqnarray}
\label{eqn:2.10}
a_s = \frac{\alpha_s}{4\pi} \,, \qquad \alpha_s = \frac{g^2}{4\pi}\,.
\end{eqnarray}
The coefficients $\beta_0$ and $z_{\xi}$, which originate from 
coupling constant and gauge constant renormalization, are given by
\begin{eqnarray}
\label{eqn:2.11}
\hat a_s = a_s \Big [ 1 + a_s S_{\varepsilon}
\Big (2\beta_0 \frac{1}{\varepsilon} \Big )\Big ] \,,
\end{eqnarray}
\begin{eqnarray}
\label{eqn:2.12}
\hat \xi = \xi \Big [ 1 + a_s S_{\varepsilon}
\Big (z_\xi\frac{1}{\varepsilon} \Big )\Big ]\,,
\end{eqnarray}
where
\begin{eqnarray}
\label{eqn:2.13}
\beta_0= \frac{11}{3} C_A - \frac{8}{3} n_f T_f\,, 
\qquad
z_{\xi}= C_A \Big ( -\frac{10}{3} - (1- \xi) \Big ) + \frac{8}{3}n_f T_f\,. 
\end{eqnarray}
In QCD ($SU(N)$) the colour factors are given by $C_F=(N^2-1)/2N$, $C_A=N$,
$T_f=1/2$ and $n_f$ stands for the number of light flavours.
Finally the $\gamma_{ij}^{(k)}$ denote the 
coefficients of the order $a_s^{k+1}$
terms appearing in the series expansions of the anomalous dimensions.
Using the same notation we can also express the unphysical part of the 
non-singlet OME in (\ref{eqn:2.5}) in the aforementioned renormalization group 
coefficients
\begin{eqnarray}
\label{eqn:2.14}
\hat A_{qq}^{{\rm NS, EOM}}&=&  \hat a_s S_{\varepsilon} 
\big(\frac{-p^2}{\mu^2}\big)
^{\varepsilon/2} \Big [  b_{qq}^{{\rm NS},(1)}
+ \varepsilon b_{qq}^{{\rm NS},\varepsilon,(1)}\Big ]
\nonumber\\[2ex]
&& +{\hat a}_s^2 S_{\varepsilon}^2 \big(\frac{-p^2}{\mu^2}\big)^{\varepsilon}
 \Big[
\frac{1}{\varepsilon} \Big \{ \gamma_{qq}^{{\rm NS},(0)} b_{qq}^{{\rm NS},(1)}
- 2 \beta_0 b_{qq}^{{\rm NS},(1)} 
\nonumber\\[2ex]
&& - \hat{\xi}  \frac{d~b_{qq}^{{\rm NS},(1)}}
{d \hat{\xi} } z_{\xi} \Big \}
+ b_{qq}^{{\rm NS},(2)} - 2 \beta_0 b_{qq}^{{\rm NS},\varepsilon,(1)}
+ \gamma_{qq}^{{\rm NS},(0)} b_{qq}^{{\rm NS},\varepsilon,(1)}
\nonumber\\[2ex]
&& - \hat{\xi}  \frac{d~b_{qq}^{{\rm NS},\varepsilon,(1)}} {d \hat{\xi} }
z_{\xi} \Big ]_{\hat{\xi}=1} \,.
\end{eqnarray}

Since the singlet OME is only computed up to second order the 
unphysical part arises from
the fact that the equations of motion are not satisfied. In this order there
is no need to introduce NGI operators. 
The singlet OME can be split into nonsinglet (NS) and purely singlet (PS) 
parts 
\begin{eqnarray}
\label{eqn:2.15}
A_{qq}^{\rm S}= A_{qq}^{\rm NS}+A_{qq}^{\rm PS}\,,
\end{eqnarray}
where the purely singlet physical OME is
\begin{eqnarray}
\label{eqn:2.16}
\hat A_{qq}^{{\rm PS, PHYS}} &=&
 {\hat a}_s^2 S_{\varepsilon}^2 
\big(\frac{-p^2}{\mu^2}\big)^{\varepsilon} \Big[
 \frac{1}{\varepsilon^2} \Big \{ \frac{1}{2} \gamma_{qg}^{(0)}
\gamma_{gq}^{(0)} \Big \}
\nonumber\\[2ex]
&& + \frac{1}{\varepsilon}\Big \{ \frac{1}{2} \gamma_{qq}^{{\rm PS},(1)}
+ \gamma_{qg}^{(0)} a_{gq}^{(1)} \Big \}
+ a_{qq}^{{\rm PS},(2)} + \gamma_{qg}^{(0)} a_{gq}^{\varepsilon,(1)}\Big ]\,,
\end{eqnarray}
and the purely singlet unphysical OME is
\begin{eqnarray}
\label{eqn:2.17}
\hat A_{qq}^{{\rm PS, EOM}}&=&
{\hat a}_s^2 S_{\varepsilon}^2 (\frac{-p^2}{\mu^2})^{\varepsilon} \Big[
\frac{1}{\varepsilon} \Big \{ \gamma_{qg}^{(0)} b_{gq}^{(1)} \Big \}
+ b_{qq}^{{\rm PS},(2)} + \gamma_{qg}^{(0)} b_{gq}^{\varepsilon,(1)} \Big ]\,.
\end{eqnarray}

The mixing with NGI operators does occur 
in $\hat A_{qg,\mu\nu}$ which can be written as
\begin{eqnarray}
\label{eqn:2.18}
\hat A_{qg,\mu\nu}
=\hat A_{qg}^{\rm PHYS} \, T_{\mu\nu}^{(1)} +
\hat A_{qg}^{\rm EOM} \, T_{\mu\nu}^{(2)}+ \hat A_{qg}^{\rm NGI} \,
T_{\mu\nu}^{(3)} \,. 
\end{eqnarray}
In this expression the tensors are given by
\begin{eqnarray}
\label{eqn:2.19}
T_{\mu\nu}^{(1)} = \frac{1 + (-1)^n}{2} \Big [ g_{\mu\nu} - \frac{p_{\mu}
\Delta_{\nu}+\Delta_{\mu}p_{\nu} }{\Delta \cdot p} 
+ \frac{\Delta_{\mu}\Delta_{\nu}
p^2}{(\Delta \cdot p)^2} \Big ] (\Delta \cdot p)^n \,,
\end{eqnarray}
\begin{eqnarray}
\label{eqn:2.20}
T_{\mu\nu}^{(2)} = \frac{1 + (-1)^n}{2} \Big [ \frac{p_{\mu}p_{\nu}}{p^2}
- \frac{p_{\mu}\Delta_{\nu}+\Delta_{\mu}p_{\nu} }{\Delta \cdot p} 
+ \frac{\Delta_{\mu}\Delta_{\nu}
p^2}{(\Delta \cdot p)^2} \Big ] (\Delta \cdot p)^n \,,
\end{eqnarray}
and
\begin{eqnarray}
\label{eqn:2.21}
T_{\mu\nu}^{(3)} = \frac{1 + (-1)^n}{2} \Big [ - \frac{p_{\mu}
\Delta_{\nu}+\Delta_{\mu}p_{\nu} }{\Delta \cdot p} 
+ 2 \frac{\Delta_{\mu}\Delta_{\nu}
p^2}{(\Delta \cdot p)^2} \Big ] (\Delta \cdot p)^n \,.
\end{eqnarray}
For later purposes we also define the tensor
\begin{eqnarray}
\label{eqn:2.22}
T_{\mu\nu}^{(4)} = \frac{1 + (-1)^n}{2} \Big [  \frac{p_{\mu}
\Delta_{\nu}+\Delta_{\mu}p_{\nu} }{\Delta \cdot p} 
+ 2 \frac{\Delta_{\mu}\Delta_{\nu}
p^2}{(\Delta \cdot p)^2} \Big ] (\Delta \cdot p)^n \,.
\end{eqnarray}
The above tensors satisfy the following relations
\begin{eqnarray}
\label{eqn:2.23}
p^{\mu}\,T_{\mu\nu}^{(i)}=0 \qquad (i=1,2) \,,
\qquad p^{\mu}\,T_{\mu\nu}^{(i)} \not = 0
\qquad (i=3,4) \,,
\end{eqnarray}
\begin{eqnarray}
\label{eqn:2.24}
p^{\mu}p^{\nu}\,T_{\mu\nu}^{(i)}=0 \qquad (i=1,2,3) \,, \qquad 
p^{\mu}p^{\nu}\,T_{\mu\nu}^{(4)} \not = 0 \,.
\end{eqnarray}
Using these relations one can show that the OME's in Eq. (\ref{eqn:2.18})
satisfy the following Ward-identities (WI) for $i=q$
\begin{eqnarray}
\label{eqn:2.25}
p^{\mu}\hat A_{ig,\mu\nu} = \frac{1 + (-1)^n}{2} \Big [ -p_{\nu} +
\frac{\Delta_{\nu} p^2}{\Delta \cdot p} \Big ] (\Delta \cdot p)^n 
\hat A_{ig}^{\rm NGI} \,,
\end{eqnarray}
\begin{eqnarray}
\label{eqn:2.26}
p^{\mu}p^{\nu} \hat A_{ig,\mu\nu} = 0 \,.
\end{eqnarray}
The WI in Eq. (\ref{eqn:2.25}) shows that the unphysical part 
in Eq. (\ref{eqn:2.18}), given by $\hat A_{qg}^{\rm NGI}$, is due to the 
NGI operator $O_A$ in Eq. (\ref{eqn:B1}). 
The second term in Eq. (\ref{eqn:2.18}), given by
$\hat A_{qg}^{\rm EOM}$, is also unphysical  
due to the fact that we cannot apply the equations of motion.
The physical part of the OME in Eq. (\ref{eqn:2.18}) becomes
\begin{eqnarray}
\label{eqn:2.27}
\hat A_{qg}^{\rm PHYS}&=&  \hat a_s S_{\varepsilon} 
\big(\frac{-p^2}{\mu^2}\big)
^{\varepsilon/2} \Big [ \frac{1}{\varepsilon} \gamma_{qg}^{(0)} + a_{qg}^{(1)}
+ \varepsilon a_{qg}^{\varepsilon,(1)}\Big ]
\nonumber\\[2ex]
&&+{\hat a}_s^2 S_{\varepsilon}^2 
\big(\frac{-p^2}{\mu^2}\big)^{\varepsilon} \Big[
 \frac{1}{\varepsilon^2} \Big \{ \frac{1}{2} \Big (\gamma_{qq}^{(0)} 
 \gamma_{qg}^{(0)} + \gamma_{qg}^{(0)} \gamma_{gg}^{(0)} \Big )
- \beta_0 \gamma_{qg}^{(0)} \Big \}
\nonumber\\[2ex]
&& +\frac{1}{\varepsilon} \Big \{ \frac{1}{2} \gamma_{qg}^{(1)}
- 2 \beta_0 a_{qg}^{(1)} +
\gamma_{qg}^{(0)} a_{gg}^{(1)} + a_{qg}^{(1)} \gamma_{qq}^{(0)}
- \hat{\xi}  \frac{d~a_{qg}^{(1)}}
{d \hat{\xi} } z_{\xi} \Big \}
\nonumber\\[2ex]
&& + a_{qg}^{(2)}  - 2 \beta_0 a_{qg}^{\varepsilon,(1)}
+ \gamma_{qg}^{(0)} a_{gg}^{\varepsilon,(1)} + \gamma_{qq}^{(0)} 
a_{qg}^{\varepsilon,(1)}
\nonumber\\[2ex]
&& - \hat{\xi}  \frac{d~a_{qg}^{\varepsilon,(1)}} {d \hat{\xi} } z_{\xi}
\Big ]_{\hat{\xi}=1} \,.
\end{eqnarray}
The unphysical parts are given by
\begin{eqnarray}
\label{eqn:2.28}
\hat A_{qg}^{\rm EOM}&=&  \hat a_s S_{\varepsilon} 
\big(\frac{-p^2}{\mu^2}\big)
^{\varepsilon/2} \Big [  b_{qg}^{(1)}
+ \varepsilon b_{qg}^{\varepsilon,(1)}\Big ]
\nonumber\\[2ex]
&& +{\hat a}_s^2 S_{\varepsilon}^2 
\big(\frac{-p^2}{\mu^2}\big)^{\varepsilon} \Big[
\frac{1}{\varepsilon} \Big \{  \gamma_{qg}^{(0)}  b_{gg}^{(1)}
+ \gamma_{qq}^{(0)}  b_{qg}^{(1)}- 2 \beta_0 b_{qg}^{(1)}
\nonumber\\[2ex]
&& - \hat{\xi}  \frac{d~b_{qg}^{(1)}}
{d \hat{\xi} } z_{\xi} \Big \} + b_{qg}^{(2)} - 2 \beta_0 
b_{qg}^{\varepsilon,(1)}
+ \gamma_{qg}^{(0)}  b_{gg}^{\varepsilon,(1)} + \gamma_{qq}^{(0)}  
b_{qg}^{\varepsilon,(1)}
\nonumber\\[2ex]
&& - \hat{\xi}  \frac{d~b_{qg}^{\varepsilon,(1)}}{d \hat{\xi} } z_{\xi}
\Big ]_{\hat{\xi}=1}\,,
\end{eqnarray}
and
\begin{eqnarray}
\label{eqn:2.29}
\hat A_{qg}^{\rm NGI}&=&
{\hat a}_s^2 S_{\varepsilon}^2 
\big(\frac{-p^2}{\mu^2}\big)^{\varepsilon} \Big[
 \frac{1}{\varepsilon^2} \Big \{ \gamma_{qg}^{(0)}\gamma_{gA}^{(0)} \Big \}
+\frac{1}{\varepsilon} \Big \{  \gamma_{qg}^{(0)} a_{gA}^{(1)}
 \Big \} + a_{qA}^{(2)}
\nonumber\\[2ex]
&& + \gamma_{qg}^{(0)} a_{gA}^{\varepsilon,(1)} \Big ] \,.
\end{eqnarray}
The coefficients with a subscript $A$ originate from the NGI operator
$O_A$ whose matrix element will be presented below when we discuss the 
renormalization. 

The next OME $\hat A_{gq}$ can be decomposed as
\begin{eqnarray}
\label{eqn:2.30}
\hat A_{gq}= \Big [
{\Delta \hspace{-0.52em}/\hspace{0.1em}}
%\Ds 
\hat A_{gq}^{\rm PHYS} + 
{p \hspace{-0.52em}/\hspace{0.1em}}
%\ps 
\frac{\Delta\cdot p}{p^2} 
\hat A_{gq}^{\rm EOM}\Big ](\Delta\cdot p)^n\,,
\end{eqnarray}
where the last term again represents the unphysical part due to the 
breakdown of the equations of motion. The physical part equals
\begin{eqnarray}
\label{eqn:2.31}
\hat A_{gq}^{\rm PHYS} &=& \hat a_s S_{\varepsilon} (\frac{-p^2}{\mu^2})
^{\varepsilon/2}
\Big [ \frac{1}{\varepsilon} \gamma_{gq}^{(0)} + a_{gq}^{(1)}
+ \varepsilon a_{gq}^{\varepsilon,(1)}  \Big ]
\nonumber\\[2ex]
&& +{\hat a}_s^2 S_{\varepsilon}^2 (\frac{-p^2}{\mu^2})^{\varepsilon} \Big[
 \frac{1}{\varepsilon^2} \Big \{ \frac{1}{2} \gamma_{gq}^{(0)}
( \gamma_{gg}^{(0)} + \gamma_{qq}^{{\rm NS},(0)} )
 - \beta_0 \gamma_{gq}^{(0)} \Big \}
\nonumber\\[2ex]
&& + \frac{1}{\varepsilon}\Big \{ \frac{1}{2} \gamma_{gq}^{(1)}
- 2 \beta_0 a_{gq}^{(1)} +
\gamma_{gq}^{(0)} a_{qq}^{{\rm NS},(1)} + \gamma_{gg}^{(0)} a_{gq}^{(1)} +
\gamma_{gA}^{(0)} (a_{Aq}^{(1)} + a_{Bq}^{(1)} )
\nonumber\\[2ex]
&& - \hat{\xi}  \frac{d~a_{gq}^{(1)}}
{d \hat{\xi} } z_{\xi} \Big \} + a_{gq}^{(2)}
-2 \beta_0 a_{gq}^{\varepsilon,(1)} + \gamma_{gq}^{(0)} 
a_{qq}^{{\rm NS},\varepsilon, (1)}
+ \gamma_{gg}^{(1)} a_{gq}^{\varepsilon, (1)}
\nonumber\\[2ex]
&& + \gamma_{gA}^{(0)} (a_{Aq}^{\varepsilon,(1)}
+ a_{Bq}^{\varepsilon,(1)} )
- \hat{\xi}  \frac{d~a_{gq}^{\varepsilon,(1)}}{d \hat{\xi} } z_{\xi}
\Big ]_{\hat{\xi}=1} \,.
\end{eqnarray}
The coefficients with the subscripts $A$ 
and $B$ refer to the NGI operators
$O_A$ and $O_B$ presented in Eq. (\ref{eqn:B1}) and Eq. (\ref{eqn:B5})
respectively.
The latter shows up for the first time in 
order $\alpha_s^2$ in the physical part of $\hat A_{gq}$. The 
corresponding OME's will be given below when we
discuss the renormalization of the physical operators. 
The unphysical part of Eq. (\ref{eqn:2.30}) equals
\begin{eqnarray}
\label{eqn:2.32}
\hat A_{gq}^{\rm EOM}&=&  \hat a_s S_{\varepsilon} 
\big(\frac{-p^2}{\mu^2}\big)
^{\varepsilon/2} \Big [  b_{gq}^{(1)}
+ \varepsilon b_{gq}^{\varepsilon,(1)}\Big ]
\nonumber\\[2ex]
&& +{\hat a}_s^2 S_{\varepsilon}^2 
\big(\frac{-p^2}{\mu^2}\big)^{\varepsilon} \Big[
\frac{1}{\varepsilon} \Big \{  \gamma_{gq}^{(0)}  b_{qq}^{(1)}
+ \gamma_{gg}^{(0)}  b_{gq}^{(1)}- 2 \beta_0 b_{gq}^{(1)} 
\nonumber\\[2ex]
&& - \hat{\xi}  \frac{d~b_{gq}^{(1)}}
{d \hat{\xi} } z_{\xi} \Big \} + b_{gq}^{(2)} -2 \beta_0 
b_{gq}^{\varepsilon,(1)}
+ \gamma_{gq}^{(0)}  b_{qq}^{{\rm NS},\varepsilon,(1)}+ \gamma_{gg}^{(0)}  
b_{gq}^{\varepsilon,(1)}
\nonumber\\[2ex]
&& - \hat{\xi}  \frac{d~b_{gq}^{\varepsilon,(1)}}{d \hat{\xi} } z_{\xi}
\Big ]_{\hat{\xi}=1}\,.
\end{eqnarray}
The last OME $\hat A_{gg,\mu\nu}$ has a similar decomposition to the 
one presented in Eq. (\ref{eqn:2.18})
\begin{eqnarray}
\label{eqn:2.33}
\hat A_{gg,\mu\nu}=\hat A_{gg}^{\rm PHYS}\, T_{\mu\nu}^{(1)} +
\hat A_{gg}^{\rm EOM} \, T_{\mu\nu}^{(2)}
+ \hat A_{gg}^{\rm NGI} \, T_{\mu\nu}^{(3)} \,.
\end{eqnarray}
In particular it satisfies the same Ward-identities as listed in
Eq. (\ref{eqn:2.25}) and Eq. (\ref{eqn:2.26}) for $i=g$. 
Hence the second (unphysical) part in this expression 
originates from the breakdown of the equations 
of motion whereas the third term again originates from the NGI operator
$O_A$ in Eq. (B1). The physical part of the OME is equal to
\begin{eqnarray}
\label{eqn:2.34}
\hat A_{gg}^{\rm PHYS}&=& 1 + \hat a_s S_{\varepsilon} 
\big(\frac{-p^2}{\mu^2}\big)
^{\varepsilon/2} \Big [ \frac{1}{\varepsilon} \gamma_{gg}^{(0)} + a_{gg}^{(1)}
+ \varepsilon a_{gg}^{\varepsilon,(1)}\Big ]
\nonumber\\[2ex]
&&+{\hat a}_s^2 S_{\varepsilon}^2 
\big(\frac{-p^2}{\mu^2}\big)^{\varepsilon} \Big[
 \frac{1}{\varepsilon^2} \Big \{ \frac{1}{2} \Big ((\gamma_{gg}^{(0)})^2
+ \gamma_{gq}^{(0)} \gamma_{qg}^{(0)} \Big )
- \beta_0 \gamma_{gg}^{(0)} \Big \}
\nonumber\\[2ex]
&& +\frac{1}{\varepsilon} \Big \{ \frac{1}{2} \gamma_{gg}^{(1)}
- 2 \beta_0 a_{gg}^{(1)} +
\gamma_{gg}^{(0)} a_{gg}^{(1)} + \gamma_{gq}^{(0)} a_{qg}^{(1)}
+ \gamma_{gA}^{(0)} (a_{Ag}^{(1)} + a_{\omega g}^{(1)} )
\nonumber\\[2ex]
&& - \hat{\xi}  \frac{d~a_{gg}^{(1)}}
{d \hat{\xi} } z_{\xi} \Big \} + a_{gg}^{(2)} - 2 \beta_0 
a_{gg}^{\varepsilon,(1)}
+ \gamma_{gg}^{(0)} a_{gg}^{\varepsilon,(1)} + \gamma_{gq}^{(0)} 
a_{qg}^{\varepsilon,(1)}
\nonumber\\[2ex]
&& + \gamma_{gA}^{(0)} (a_{Ag}^{\varepsilon,(1)} + 
a_{\omega g}^{\varepsilon,(1)} )
- \hat{\xi}  \frac{d~a_{gg}^{\varepsilon,(1)}}{d \hat{\xi} } z_{\xi}
\Big ]_{\hat{\xi}=1}\,.
\end{eqnarray}
The unphysical parts are given by
\begin{eqnarray}
\label{eqn:2.35}
\hat A_{gg}^{\rm EOM}&=&  \hat a_s S_{\varepsilon} 
\big(\frac{-p^2}{\mu^2}\big)
^{\varepsilon/2} \Big [  b_{gg}^{(1)}
+ \varepsilon b_{gg}^{\varepsilon,(1)}\Big ]
\nonumber\\[2ex]
&& +{\hat a}_s^2 S_{\varepsilon}^2 (\frac{-p^2}{\mu^2})^{\varepsilon} \Big[
\frac{1}{\varepsilon} \Big \{  \gamma_{gg}^{(0)}  b_{gg}^{(1)}
- 2 \beta_0 b_{gg}^{(1)}
+ \gamma_{gA}^{(0)}(b_{Ag}^{(1)} + b_{\omega g}^{(1)} )
\nonumber\\[2ex]
&& - \hat{\xi}  \frac{d~b_{gg}^{(1)}}
{d \hat{\xi} } z_{\xi} \Big \} + b_{gg}^{(2)} - 2 \beta_0 
b_{gg}^{\varepsilon,(1)}
+ \gamma_{gA}^{(0)}(b_{Ag}^{\varepsilon,(1)} + b_{\omega g}^{\varepsilon,(1)} )
\nonumber\\[2ex]
&& - \hat{\xi}  \frac{d~b_{gg}^{\varepsilon,(1)}}{d \hat{\xi} } z_{\xi}
\Big ]_{\hat{\xi}=1}\,,
\end{eqnarray}
and
\begin{eqnarray}
\label{eqn:2.36}
\hat A_{gg}^{\rm NGI}&=&  \hat a_s S_{\varepsilon} 
\big(\frac{-p^2}{\mu^2}\big)
^{\varepsilon/2} \Big [ \frac{1}{\varepsilon} \gamma_{gA}^{(0)} + a_{gA}^{(1)}
+ \varepsilon a_{gA}^{\varepsilon,(1)}\Big ]
\nonumber\\[2ex]
&& +{\hat a}_s^2 S_{\varepsilon}^2 
\big(\frac{-p^2}{\mu^2}\big)^{\varepsilon} \Big[
 \frac{1}{\varepsilon^2} \Big \{ \frac{1}{2} \Big (\gamma_{gg}^{(0)}
+ \gamma_{Ag}^{(0)} +\gamma_{\omega g}^{(0)} -2\beta_0
\Big )\gamma_{gA}^{(0)} \Big \}
\nonumber\\[2ex]
&& +\frac{1}{\varepsilon} \Big \{ \frac{1}{2} \gamma_{gA}^{(1)}
- 2 \beta_0 a_{gA}^{(1)} +
\gamma_{gg}^{(0)} a_{gA}^{(1)}   + \gamma_{gA}^{(0)}
(a_{AA}^{(1)} + a_{\omega A}^{(1)} )
 - \hat{\xi}  \frac{d~a_{gA}^{(1)}}
{d \hat{\xi} } z_{\xi} \Big \} 
\nonumber\\[2ex]
&& + a_{gA}^{(2)} - 2 \beta_0 
a_{gA}^{\varepsilon,(1)}
+ \gamma_{gg}^{(0)} a_{gA}^{\varepsilon,(1)} + \gamma_{gA}^{(0)} 
(a_{AA}^{\varepsilon,(1)}+ a_{\omega A}^{\varepsilon,(1)} )
\nonumber\\[2ex]
&& - \hat{\xi}  \frac{d~a_{gA}^{\varepsilon,(1)}}{d \hat{\xi} } z_{\xi}
\Big ]_{\hat{\xi}=1}\,,
\end{eqnarray}
respectively. In these expressions we observe that there are quantities
with the subscript $A$ and with the subscript $\omega$.
The latter originate from the NGI ghost operator $O_{\omega}$
presented in Eq. (\ref{eqn:B2}).
The renormalization of the above OME's involves the mixing of the NGI
(non gauge invariant) operators $O_A$, $O_B$ and $O_{\omega}$ mentioned above 
with the physical (gauge invariant) operators 
in Eqs. (\ref{eqn:2.1})-(\ref{eqn:2.3}). 
Therefore we have to compute the matrix elements in 
Eq. (\ref{eqn:2.5}) where the physical operators indicated by $i=q,g$ are 
replaced by the NGI ones labelled by $i=A,B,\omega$.
To get the physical OME's which are
finite up to order $\alpha_s^2$ the unphysical ones have 
fortunately only to be calculated up to order $\alpha_s$.

The operator renormalization proceeds as follows.
First we have to perform coupling constant and gauge constant renormalization.
This is achieved by replacing the bare constants by the renormalized ones
by substituting Eqs. (\ref{eqn:2.11}) and (\ref{eqn:2.12}) in the above
expressions for the OME's. Subsequently the OME's have to be multiplied
by the operator renormalization constants to remove the remaining ultraviolet
divergences. The most simple case is the renormalization of the non-singlet
OME's since here we do not have mixing with physical and NGI operators.
This renormalization is achieved by
\begin{eqnarray}
\label{eqn:2.37}
A_{qq}^{{\rm NS}, i} = (Z^{-1})_{qq}^{\rm NS} \hat A_{qq}^{{\rm NS}, i}\,,
\end{eqnarray}
with $i=$ PHYS and EOM. The inverse of the operator renormalization 
constant equals
\begin{eqnarray}
\label{eqn:2.38}
(Z^{-1})_{qq}^{\rm NS} &=& 1 
+ a_s S_{\varepsilon} \Big [- \frac{1}{\varepsilon}
\gamma_{qq}^{{\rm NS},(0)} \Big ]
\nonumber\\[2ex]
&& + a_s^2 S_{\varepsilon}^2 \Big [ \frac{1}{\varepsilon^2} \Big \{\frac{1}{2}
(\gamma_{qq}^{{\rm NS},(0)})^2
- \beta_0 \gamma_{qq}^{{\rm NS},(0)} \Big \} -  \frac{1}{2\varepsilon}
\gamma_{qq}^{{\rm NS},(1)}  \Big ]\,.
\end{eqnarray}
In the case of the singlet operators $A_{qq}^{{\rm S},i}$ there is no 
mixing between GI and NGI operators (at least up to order $\alpha_s^2$).
The singlet OME's are finite for $i=$ PHYS and EOM when
\begin{eqnarray}
\label{eqn:2.39}
A_{qq}^{{\rm S}, i}= (Z^{-1})_{qq}^{\rm S} \hat A_{qq}^{{\rm S}, i}
+ (Z^{-1})_{qg} \hat A_{gq}^{i}\,,
\end{eqnarray}
and the renormalization involves the matrix
$Z_{ij}$ due to mixing with unphysical operators. Hence we have to invert
this matrix in order to perform the operator renormalization. The first 
renormalization constant in Eq. (\ref{eqn:2.39}) can be split as follows
\begin{eqnarray}
\label{eqn:2.40}
(Z^{-1})_{qq}^{\rm S}= (Z^{-1})_{qq}^{\rm NS}+ (Z^{-1})_{qq}^{\rm PS} \,,
\end{eqnarray}
where the first part is equal to the expression (\ref{eqn:2.38}) and the last
part equals
\begin{eqnarray}
\label{eqn:2.41}
(Z^{-1})_{qq}^{\rm PS}&=&
 a_s^2 S_{\varepsilon}^2 \Big [ \frac{1}{\varepsilon^2} \Big \{\frac{1}{2}
\gamma_{qg}^{(0)} \gamma_{gq}^{(0)} \Big \} -  \frac{1}{2\varepsilon}
\gamma_{qq}^{{\rm PS},(1)}  \Big ]\,.
\end{eqnarray}
The constant $(Z^{-1})_{qg}$ in Eq. (\ref{eqn:2.39}) also shows up in the
renormalization of the next OME, namely
\begin{eqnarray}
\label{eqn:2.42}
A_{qg}^{i}= (Z^{-1})_{qq}^{\rm S} \hat A_{qg}^{i}
+ (Z^{-1})_{qg} \hat A_{gg}^{i}\,,
\end{eqnarray}
for $i=$ PHYS, EOM and NGI.
It can be decomposed into renormalization group coefficients as follows
\begin{eqnarray}
\label{eqn:2.43}
(Z^{-1})_{qg}&=&  a_s S_{\varepsilon} \Big [- \frac{1}{\varepsilon}
\gamma_{qg}^{(0)} \Big ]+ a_s^2 S_{\varepsilon}^2 \Big[
\frac{1}{\varepsilon^2} \Big \{ \frac{1}{2} \gamma_{qg}^{(0)}(\gamma_{qq}^{(0)}
+ \gamma_{gg}^{(0)} ) - \beta_0 \gamma_{qg}^{(0)}\Big \}
\nonumber\\[2ex]
&& -\frac{1}{2\varepsilon} \gamma_{qg}^{(1)} \Big]\,.
\end{eqnarray}
For the renormalization of $A_{gq}^{\rm PHYS}$ in Eq. (\ref{eqn:2.31}) and 
$A_{gq}^{\rm EOM}$ in Eq. (\ref{eqn:2.32})
we also need the contributions of the NGI OME's. Here the finite
expressions are given by
\begin{eqnarray}
\label{eqn:2.44}
A_{gq}^{i}= (Z^{-1})_{gg} \Big [\hat A_{gq}^{i}+ \eta \Big (
\hat A_{Aq}^{i} +  \hat A_{Bq}^{i} \Big ) \Big ]
+ (Z^{-1})_{gq} \hat A_{qq}^{{\rm S},i} \,,
\end{eqnarray}
where $\eta$ is defined as a source multiplying the NGI operators
in the effective action (see Eq. (2.8) in \cite{hane}). 
Here we will write this quantity as follows
\begin{eqnarray}
\label{eqn:2.45}
\eta &=&a_s S_{\varepsilon} \Big [-\frac{1}{\varepsilon}\gamma_{gA}^{(0)} 
\Big ]
+ a_s^2 S_{\varepsilon}^2 \Big [\frac{1}{\varepsilon^2} \Big \{ \frac{1}{2}
\Big (\gamma_{AA}^{(0)}+\gamma_{\omega A}^{(0)}-\gamma_{gg}^{(0)}
\nonumber\\[2ex]
&& - 2 \beta_0  \Big )\gamma_{gA}^{(0)} \Big \}
- \frac{1}{2 \varepsilon}\gamma_{gA}^{(1)} \Big ]\,.
\end{eqnarray}
If we sandwich the operator $O_A$ between quark states (see 
Eq. (\ref{eqn:2.5})) we get 
\begin{eqnarray}
\label{eqn:2.46}
\hat A_{Aq}^{\rm PHYS}
=\hat a_s S_{\varepsilon} \big(\frac{-p^2}{\mu^2}\big)^{\varepsilon/2}
\Big [ \frac{1}{\varepsilon} \gamma_{Aq}^{(0)} + a_{Aq}^{(1)}
+ \varepsilon a_{Aq}^{\varepsilon,(1)}\Big ]\,,
\end{eqnarray}
and
\begin{eqnarray}
\label{eqn:2.47}
\hat A_{Aq}^{\rm EOM}=\hat a_s S_{\varepsilon} 
\big(\frac{-p^2}{\mu^2}\big)^{\varepsilon/2}
\Big [  b_{Aq}^{(1)} + \varepsilon b_{Aq}^{\varepsilon,(1)}\Big ]\,,
\end{eqnarray}
up to lowest order.
We proceed in a similar way for the operator $O_B$ 
and the OME's read
\begin{eqnarray}
\label{eqn:2.48}
\hat A_{Bq}^{\rm PHYS}=\hat a_s S_{\varepsilon} 
\big(\frac{-p^2}{\mu^2}\big)^{\varepsilon/2}
\Big [ - \frac{1}{\varepsilon} \gamma_{Aq}^{(0)} + a_{Bq}^{(1)}
+ \varepsilon a_{Bq}^{\varepsilon,(1)}\Big ]\,,
\end{eqnarray}
and 
\begin{eqnarray}
\label{eqn:2.49}
\hat A_{Bq}^{\rm EOM}=\hat a_s S_{\varepsilon} 
\big(\frac{-p^2}{\mu^2}\big)^{\varepsilon/2}
\Big [  b_{Bq}^{(1)} + \varepsilon b_{Bq}^{\varepsilon,(1)}\Big ]\,.
\end{eqnarray}
Notice that $\hat A_{Aq}$ and $\hat A_{Bq}$ can be 
decomposed in a similar way to
$\hat A_{gq}$ in Eq. (\ref{eqn:2.30}). The operator renormalization constant
appearing in the last part of Eq. (\ref{eqn:2.44}) can now be written as
\begin{eqnarray}
\label{eqn:2.50}
(Z^{-1})_{gq}&=&  a_s S_{\varepsilon} \Big [- \frac{1}{\varepsilon}
\gamma_{gq}^{(0)} \Big ]+ a_s^2 S_{\varepsilon}^2 \Big[
\frac{1}{\varepsilon^2} \Big \{ \frac{1}{2} \gamma_{gq}^{(0)}(\gamma_{gg}^{(0)}
+ \gamma_{qq}^{(0)} ) - \beta_0 \gamma_{gq}^{(0)}\Big \}
\nonumber\\[2ex]
&& -\frac{1}{2\varepsilon} \gamma_{gq}^{(1)} \Big]\,.
\end{eqnarray}
The first constant $(Z^{-1})_{gg}$ in Eq. (\ref{eqn:2.44}) 
also shows up in the renormalization of the OME
\begin{eqnarray}
\label{eqn:2.51}
A_{gg}^{i}= (Z^{-1})_{gq} \hat A_{qg}^{i}
+ (Z^{-1})_{gg} \Big [\hat A_{gg}^{i} + \eta \Big ( \hat A_{Ag}^{i}+
\hat A_{\omega g}^{i} \Big ) \Big] \,,
\end{eqnarray}
for $i=$ PHYS, EOM and NGI.
If we express this constant into the renormalization group 
coefficients in the following way
\begin{eqnarray}
\label{eqn:2.52}
(Z^{-1})_{gg}&=& 1 + a_s S_{\varepsilon} \Big [- \frac{1}{\varepsilon}
\gamma_{gg}^{(0)} \Big ]
\nonumber\\[2ex]
&& + a_s^2 S_{\varepsilon}^2 \Big [ \frac{1}{\varepsilon^2} \Big \{\frac{1}{2}
(\gamma_{gg}^{(0)})^2 + \frac{1}{2} \gamma_{gq}^{(0)}\gamma_{qg}^{(0)}
- \beta_0 \gamma_{gg}^{(0)} \Big \} -  \frac{1}{2\varepsilon}
\gamma_{gg}^{(1)}  \Big ]\,,
\end{eqnarray}
we obtain the finite physical (PHYS) and unphysical parts (EOM, NGI) 
of expression (\ref{eqn:2.51}). In order to obtain these finite parts one
has first to calculate the OME's which emerge when the NGI operators
$O_A$ and $O_{\omega}$ are sandwiched between gluon states (see 
Eq. (\ref{eqn:2.5})). They can be decomposed as follows
\begin{eqnarray}
\label{eqn:2.53}
\hat A_{Ag,\mu\nu}=\hat A_{Ag}^{\rm PHYS} \, T_{\mu\nu}^{(1)} 
+ \hat A_{Ag}^{\rm EOM}\,T_{\mu\nu}^{(2)}
+ \hat A_{Ag}^{\rm NGI}\, T_{\mu\nu}^{(3)} +
\hat A_{Ag}^{\rm WI} \,T_{\mu\nu}^{(4)}\,,
\end{eqnarray}
with
\begin{eqnarray}
\label{eqn:2.54}
\hat A_{Ag}^{\rm PHYS}&=&  \hat a_s S_{\varepsilon} 
\big(\frac{-p^2}{\mu^2}\big)
^{\varepsilon/2} \Big [ \frac{1}{\varepsilon} \gamma_{Ag}^{(0)} + a_{Ag}^{(1)}
+ \varepsilon a_{Ag}^{\varepsilon,(1)}\Big ]\,,
\end{eqnarray}
\begin{eqnarray}
\label{eqn:2.55}
\hat A_{Ag}^{\rm EOM}&=&  \hat a_s S_{\varepsilon} 
\big(\frac{-p^2}{\mu^2}\big)
^{\varepsilon/2} \Big [  b_{Ag}^{(1)}
+ \varepsilon b_{Ag}^{\varepsilon,(1)}\Big ]\,,
\end{eqnarray}
\begin{eqnarray}
\label{eqn:2.56}
\hat A_{Ag}^{\rm NGI}&=& 1 + \hat a_s S_{\varepsilon} 
\big(\frac{-p^2}{\mu^2}\big)
^{\varepsilon/2} \Big [ \frac{1}{\varepsilon} \gamma_{AA}^{(0)} + a_{AA}^{(1)}
+ \varepsilon a_{AA}^{\varepsilon,(1)}\Big ]\,,
\end{eqnarray}
and
\begin{eqnarray}
\label{eqn:2.57}
\hat A_{Ag}^{\rm WI}&=&  \hat a_s S_{\varepsilon} 
\big(\frac{-p^2}{\mu^2}\big)
^{\varepsilon/2} \Big [  c_{Ag}^{(1)}
+ \varepsilon c_{Ag}^{\varepsilon,(1)}\Big ]\,.
\end{eqnarray}
Similar algebraic expressions are obtained for $O_{\omega}$. They are given by
\begin{eqnarray}
\label{eqn:2.58}
\hat A_{\omega g,\mu\nu}=\hat A_{\omega g}^{\rm PHYS} \,T_{\mu\nu}^{(1)}
+ \hat A_{\omega g}^{\rm EOM}\, T_{\mu\nu}^{(2)}
+ \hat A_{\omega g}^{\rm NGI}\, T_{\mu\nu}^{(3)}
+ \hat A_{\omega g}^{\rm WI}\, T_{\mu\nu}^{(4)}\,,
\end{eqnarray}
with
\begin{eqnarray}
\label{eqn:2.59}
\hat A_{\omega g}^{\rm PHYS}&=& \hat a_s S_{\varepsilon} (\frac{-p^2}{\mu^2})
^{\varepsilon/2} \Big [- \frac{1}{\varepsilon} \gamma_{Ag}^{(0)}
+ a_{\omega g}^{(1)}  + \varepsilon a_{\omega g}^{\varepsilon,(1)}\Big ]\,,
\end{eqnarray}
\begin{eqnarray}
\label{eqn:2.60}
\hat A_{\omega g}^{\rm EOM}&=&  \hat a_s S_{\varepsilon} (\frac{-p^2}{\mu^2})
^{\varepsilon/2} \Big [  b_{\omega g}^{(1)}
+ \varepsilon b_{\omega g}^{\varepsilon,(1)}\Big ]\,,
\end{eqnarray}
\begin{eqnarray}
\label{eqn:2.61}
\hat A_{\omega g}^{\rm NGI}&=&  \hat a_s S_{\varepsilon} (\frac{-p^2}{\mu^2})
^{\varepsilon/2} \Big [ \frac{1}{\varepsilon} \gamma_{\omega A}^{(0)}
+ a_{\omega A}^{(1)}+ \varepsilon a_{\omega A}^{\varepsilon,(1)}\Big ]\,,
\end{eqnarray}
and
\begin{eqnarray}
\label{eqn:2.62}
\hat A_{\omega g}^{\rm WI}&=&  \hat a_s S_{\varepsilon} (\frac{-p^2}{\mu^2})
^{\varepsilon/2} \Big [ - c_{Ag}^{(1)} -  \varepsilon c_{Ag}^{\varepsilon,(1)}
\Big ]\,.
\end{eqnarray}
Notice that neither $\hat A_{Ag,\mu\nu}$ in Eq. (\ref{eqn:2.54}) nor 
$\hat A_{\omega g,\mu\nu}$ in Eq. (\ref{eqn:2.59}) satisfy the Ward-identities 
(WI) in Eq. (\ref{eqn:2.25}) and Eq. (\ref{eqn:2.26}). However if we add them
according to Eq. (\ref{eqn:2.51}) the terms proportional to the tensor
$T_{\mu\nu}^{(4)}$ cancel and the Ward-identies are restored. 

From the residues of the single pole terms appearing in the physical operator
renormalization constants $Z_{ij}$ with $i,j=q,g$ given above one 
can now read off the anomalous dimensions presented in the 
${\overline {\rm MS}}$-scheme.
Here we agree with the  results published in \cite{fupe} (see also 
\cite{hane}). Therefore we have a check that the residues of the 
single and double pole terms are correct.
By comparison with the 
algebraic expressions given above one can obtain all renormalization 
group coefficients like $a_{ij}^{(k)}$, $a_{ij}^{\varepsilon,(1)}$ etc.. 
Explicit expressions for the physical and unphysical OME's 
can be found in Appendices A and B respectively. They contain all the
finite second order terms which survive in the 
limit $\varepsilon \rightarrow 0$. These terms have not been
calculated previously.

Note that a remarkable property is found for the non-singlet OME 
$\hat A_{qq}^{\rm NS,PHYS}$ in Eq. (\ref{eqn:2.8}) for which the 
full expression is given in Eq. (\ref{eqn:A3}).
The first moment of this OME is equal to unity up to second order 
in $\alpha_s$ provided we choose the Feynman gauge ($\hat\xi=1$). This 
result is expected for the on-shell expression as it is a check of the 
Adler sum rule \cite{adler}. However the fact that all the coefficients of the
terms in $(-p^2/\mu^2)$ are zero shows that the sum rule is true up to
order $\alpha_s^2$ for the off-shell expression 
which is not an S-matrix element.
Finally we want to comment on the use of the unrenormalized expressions
given in Appendix A for the computation of the three-loop OME's. Since the 
external quark and gluon legs are off-shell one can insert our results 
into the Feynman integrals for one-loop graphs. In this way one gets
expressions corresponding to some of the three-loop graphs. It is however
clear that the most difficult Feynman integrals belonging to the non-planar
diagrams, where all quark or gluon lines cross over, remain to be done.

%%%%%%%%%%%%%%%%%%%%%%%%% ACKNOWLEDGEMENTS %%%%%%%%%%%%%%%%%%%%%%%%%%%%%%%%%%%
%
%\newpage
\acknowledgments
This research was supported in part 
by the National Science Foundation
grant PHY-9722101.
%%%%%%%%%%%%%%%%%%%%%%%%%%%%%%%%%%%%%%%%%%%%%%%%%%%%%%%%%%%%%%%
%%%%% APPENDICES %%%%%%%%%%%%%%%%%%%%%%
%\newcommand{\mysection}{\setcounter{equation}{0}\section}
%\mysection*{Appendix A}
%\setcounter{section}{1}
\appendix 
\section{}
%\appendix{Appendix A}

In this Appendix we present full expressions for the two-loop corrected 
operator matrix elements computed from the Feynman diagrams depicted in 
\cite{frs}. The second order contributions are calculated up to 
finite terms which survive in the limit $\varepsilon \rightarrow 0$.
The OME's presented here are unrenormalized and external
self-energy corrections are included.
In these expressions definitions of the Riemann zeta-functions 
$\zeta(n)$ and the polylogarithms ${\rm Li}_n(z)$, ${\rm S}_{n,m}(z)$ can be 
found in \cite{lbmr}. Also the distributions $(1/(1-z))_+$ and
$(\ln(1-z)/(1-z))_+$ are 
written simply as $1/(1-z)$ and $\ln(1-z)/(1-z)$ respectively
to shorten the formulae.
Note that the OME's given in the text are the 
moments of the functions listed here so
%(A1)
\begin{eqnarray}
\label{eqn:A1}
A_{ij}^n = \int^1_0 \, dz z^{n-1} A_{ij}(z, \frac{-p^2}{\mu^2},
\frac{1}{\varepsilon})\,,
\end{eqnarray}
where for simplicity we have not written the moment index $n$ on the 
functions. 
Also to simplify the expressions we define the phase-space factor 
%(A2)
\begin{eqnarray}
\label{eqn:A2}
F = \frac{\hat{\alpha}_s}{4\pi} S_{\varepsilon}(\frac{-p^2}{\mu^2})
^{\varepsilon/2} \,.
\end{eqnarray}

We first split $\hat A^{\rm NS}_{qq}$ into physical 
and unphysical parts following the notation in Eq. (\ref{eqn:2.7}).
The physical part is
%(A3)
\begin{eqnarray}
\label{eqn:A3}
&&\hat A_{qq}^{\rm NS, PHYS}
\Big(z, \frac{-p^2}{\mu^2},\frac{1}{\varepsilon}\Big)
=\delta(1-z)
\nonumber \\ && \qquad
+  F \,\, C_F\Big[\frac{1}{\varepsilon}\Big\{
-4-4z+\frac{8}{1-z}+6\delta(1-z)\Big\}+4-6z
\nonumber \\ && \qquad
+2(\frac{2}{1-z}-1-z)[\ln z+\ln(1-z)]
-\delta(1-z)[7-4\zeta(2)]+\frac{1-\hat{\xi}}{1-z}
\nonumber \\ && \qquad
+\varepsilon\Big\{
1-2z
+(2-3z)[\ln z+\ln(1-z)]
+\frac{1}{2}(\frac{2}{1-z}-1-z)[\zeta(2)
\nonumber \\ && \qquad
+\{\ln z+\ln(1-z)\}^2]
+\delta(1-z)[7-\frac{3}{4}\zeta(2)-4\zeta(3)]
+\frac{1}{2}(1-\hat{\xi})
\nonumber \\ && \qquad
\times[2
-\frac{1}{1-z}
+\frac{\ln(1-z)}{1-z}
+\frac{\ln z}{1-z}+\delta(1-z)\{-2+\zeta(2)\}]
\Big\}
\Big]
\nonumber \\ && \qquad
+ F^2 \,\, \Big[
\frac{1}{\varepsilon^2}\Big\{
C_F^2[
           40
          + 8z
          - \frac{48}{1-z}
          - 2\delta(1-z)[9-16\zeta(2)]
\nonumber \\ && \qquad
          +32\ln(1-z)(1
          + z
          - \frac{2}{1-z})
       -8 \ln z(
          3
          +3z
          -\frac{4}{1-z}
          )
          ]
\nonumber \\ && \qquad
+C_AC_F[
          \frac{44}{3}(1+z-\frac{2}{1-z}) 
          - 22\delta(1-z)        
]
\nonumber \\ && \qquad
       +n_fC_FT_f[
          - \frac{16}{3}(1+z-\frac{2}{1-z})
          + 8\delta(1-z)
          ]\Big\}
\nonumber \\ && \qquad
+\frac{1}{\varepsilon}\Big\{
C_F^2[- 40
          - 4z+ \frac{56}{1-z}
          +\delta(1-z)[\frac{87}{2}-36\zeta(2)- 8\zeta(3)]
\nonumber \\ && \qquad
          -4\ln(1 - z)(1-11z+\frac{1}{1-z})
          +24\ln^2(1-z)(1+z-\frac{2}{1-z})
\nonumber \\ && \qquad
          +4\ln z(9-5z-\frac{9}{1-z})
          - 16\ln z\ln(1 - z)\frac{1}{1-z}
\nonumber \\ && \qquad
          -2\ln^2 z(7+7z-\frac{8}{1-z})
          - 8{\rm Li}_2(1 - z)(1+z)]
\nonumber \\ && \qquad
+C_AC_F[- \frac{158}{9}
          + \frac{22}{9}z+ \frac{238}{9}\frac{1}{1-z}
          +\delta(1-z)[\frac{325}{6}- \frac{44}{3}\zeta(2)
\nonumber \\ && \qquad
- 12\zeta(3)] 
          +4(1+z-\frac{2}{1-z})[\zeta(2)+\frac{11}{3}\ln(1-z)
-\frac{1}{2}\ln^2z]
\nonumber \\ && \qquad
          + \ln z(\frac{34}{3}+\frac{34}{3}z-
\frac{44}{3}\frac{1}{1-z})
]
\nonumber \\ && \qquad
+n_fC_FT_f[\frac{88}{9}
          - \frac{56}{9}z
          - \frac{56}{9}\frac{1}{1-z}
          + \delta(1-z)[-\frac{58}{3}+\frac{16}{3}\zeta(2)]
\nonumber \\ && \qquad
          - \frac{8}{3}\{\ln z+2\ln(1-z)\}(1+z-\frac{2}{1-z})]\Big\}
\nonumber \\ && \qquad
+C_F^2[\frac{188}{3}+\frac{4}{3}z-\frac{56}{1-z}
          + \zeta(2)(20-4z+ \frac{32}{3}z^2-\frac{4}{1-z})
\nonumber \\ && \qquad
          + 24\zeta(3)(1-z+2z^2-\frac{1}{1-z})
          + \delta(1-z)[-\frac{541}{8}+ \frac{97}{2}\zeta(2)
\nonumber \\ && \qquad
+54\zeta(3)-\frac{74}{5}\zeta(2)^2]
          +\ln(1 - z)(-38+38z+\frac{28}{1-z})
\nonumber \\ && \qquad
          + 8\ln(1 - z)\zeta(2)(1+z-2z^2-\frac{1}{1-z})
          - \ln^2(1-z)(13-31z
\nonumber \\ && \qquad
+\frac{6}{1-z})
          + \frac{28}{3}\ln^3(1-z)(1+z-\frac{2}{1-z})
\nonumber \\ && \qquad
          - 4\ln(1 - z){\rm Li}_2(1 - z)(3-z+4z^2)
          -\ln z(\frac{68}{3}+\frac{68}{3}z-\frac{44}{1-z})
\nonumber \\ && \qquad
          + \ln z\zeta(2)(2-14z+16z^2)
          - 2\ln z\ln(1 - z)(3-17z+\frac{10}{1-z})
\nonumber \\ && \qquad
          + 2\ln z\ln^2(1-z)(3+3z-\frac{10}{1-z})
          + \ln^2 z(11-23z-\frac{16}{3}z^2
\nonumber \\ && \qquad
-\frac{15}{1-z})
          - \ln^2 z\ln(1 - z)(6-2z+8z^2+\frac{4}{1-z})
          - \ln^3 z(5+5z
\nonumber \\ && \qquad
-\frac{16}{3}\frac{1}{1-z})
          -8 \ln z{\rm Li}_2(1 - z)(1-3z+4z^2+\frac{2}{1-z})
\nonumber \\ && \qquad
          -4 {\rm Li}_2(1 - z)(6-12z+\frac{1}{1-z})
+4{\rm Li}_3(1 - z)(5-7z+12z^2)
\nonumber \\ && \qquad
          + \frac{8}{3}(9+\frac{1}{z}+12z+4z^2)
            [\ln z\ln(1+z)+{\rm Li}_2( - z)]
          -16[\ln z{\rm Li}_2(-z)
\nonumber \\ && \qquad
-2{\rm Li}_3( - z)](1-\frac{1}{1-z})
          -8{\rm S}_{12}(1 - z)(1-7z+6z^2+\frac{4}{1-z})
]
\nonumber \\ && \qquad
+C_AC_F[
          \frac{1}{27}[941
          -580z-\frac{670}{1-z}]
          -\frac{1}{3}\zeta(2)(1-17z+16z^2
           +\frac{28}{1-z})
\nonumber \\ && \qquad
          -2\zeta(3)(10+4z+12z^2-\frac{17}{1-z})
          +\delta(1-z)[-\frac{7081}{72}+\frac{301}{18}\zeta(2)
\nonumber \\ && \qquad
  + 28\zeta(3)
             +\frac{49}{5}\zeta(2)^2]
          - \frac{2}{9}\ln(1 - z)(61+16z-\frac{119}{1-z})
          + 2\ln(1 - z)\zeta(2)
\nonumber \\ && \qquad
\times(1+3z+4z^2-\frac{5}{1-z})
          + \{\frac{22}{3}\ln^2(1-z)-\ln^3z\}(1+z-\frac{2}{1-z})
\nonumber \\ && \qquad
          -2 \ln z\zeta(2)(3+z+4z^2-\frac{5}{1-z})
  + \frac{1}{6}\ln^2 z(47+47z
            +16z^2-\frac{22}{1-z})
\nonumber \\ && \qquad
          + \frac{2}{3}\ln z\ln(1 - z)(14
            +5z-\frac{19}{1-z})
  + \frac{1}{9}\ln z(14+176z+\frac{101}{1-z})
\nonumber \\ && \qquad
          + 4\ln z{\rm Li}_2(1 - z)(2+4z^2-\frac{3}{1-z})
          -4{\rm Li}_2(1 - z)(3z-\frac{1}{1-z})
\nonumber \\ && \qquad
          -4(3+\frac{1}{3z}+4z+\frac{4}{3}z^2)
             [\ln z\ln(1+z)+{\rm Li}_2( - z)]
\nonumber \\ && \qquad
          +(1-z+4z^2-\frac{1}{1-z})[\ln^2 z\ln(1-z)
          +2\ln(1-z){\rm Li}_2(1-z)
\nonumber \\ && \qquad
-6{\rm Li}_3(1 - z)]
          +8[\ln z{\rm Li}_2(-z)-2{\rm Li}_3( - z)](1-\frac{1}{1-z})
\nonumber \\ && \qquad
          +{\rm S}_{12}(1 - z)(14- 6z+ 24z^2-\frac{14}{1-z})
]
\nonumber \\ && \qquad
+n_fC_FT_f[\frac{4}{27}[
   -22+5z +\frac{32}{1-z}]
          + \delta(1-z)[\frac{569}{18}-\frac{46}{9}\zeta(2)
\nonumber \\ && \qquad
-8\zeta(3)]
          +\frac{4}{9}\{2\ln(1-z)+\ln z\}(11-7z-\frac{7}{1-z})
\nonumber \\ && \qquad
          -\frac{2}{3}[2\zeta(2)+\{2\ln(1-z)+\ln z\}^2](1+z-\frac{2}{1-z})
]
\nonumber \\ && \qquad
-2(-1)^n(C_F^2-\frac{1}{2}C_AC_F)\Big(\frac{1}{\varepsilon}\Big\{
     - 8+8z-4\ln z(1+z)
\nonumber \\ && \qquad
     +2(-1+z+\frac{2}{1+z})[2\zeta(2)+4\ln z\ln(1+z)
           -\ln^2 z
           +4{\rm Li}_2(-z)]
\Big\}
\nonumber \\ && \qquad
          - \frac{41}{3}(1-z)
          - 8\ln(1 - z)(1-z)
          + 2\zeta(2)(3-3z+\frac{8}{3}z^2)
\nonumber \\ && \qquad
          + 8(z+\frac{1}{1+z})\ln(1+z)[\zeta(2)+\ln z\ln(1+z)
          +2{\rm Li}_2(-z)]
\nonumber \\ && \qquad
          - 4(1+z)[\ln z\ln(1 - z)+{\rm Li}_2(1 - z)+{\rm Li}_3( - z)]
\nonumber \\ && \qquad
          -\frac{1}{3}\ln z(37+25z)
          -\ln^2 z(6+\frac{8}{3}z^2)
          -2\ln^2 z\ln(1 + z)(3- z- \frac{4}{1+z})
\nonumber \\ && \qquad
          - 2\zeta(3)(3+z-\frac{2}{1+z})
          + \frac{4}{3}(\frac{1}{z}+3z+4z^2)[\ln z\ln(1+z)+{\rm Li}_2( - z)]
\nonumber \\ && \qquad
          +8{\rm S}_{12}( - z)(-1+ 3z+\frac{4}{1+z})
          +4(1-z-\frac{2}{1+z})[\frac{1}{4}\ln^3 z
\nonumber \\ && \qquad
-\ln z\ln^2(1-z)
           -2\ln z\ln(1 - z)\ln(1 + z)
           +\frac{1}{2}\ln^2 z\ln(1 - z)
\nonumber \\ && \qquad
-2\ln(1 - z){\rm Li}_2(1 - z)
           -2\ln(1 + z){\rm Li}_2(1-z)
           +\ln z{\rm Li}_2(1 - z)
\nonumber \\ && \qquad
-\ln z{\rm Li}_2( - z)+2{\rm Li}_3(1 - z)
           +{\rm S}_{12}(1 - z)+{\rm S}_{12}(z^2)]
\Big)
\Big]\,.
\end{eqnarray}
Here the factor $(-1)^n$ originates from the non-planar diagrams
(namely $k$ and $o$  in figure 2 in the singlet paper of \cite{frs}). 
It multiplies that part of the matrix element which is needed for 
the mass factorization of physical processes with two identical 
quarks in the final state.

The unphysical part is given by (see Eq. (\ref{eqn:2.14}))
%
%(A4)
\begin{eqnarray}
\label{eqn:A4}
&&\hat A_{qq}^{\rm NS, EOM}
\Big(z, \frac{-p^2}{\mu^2},\frac{1}{\varepsilon}\Big)
= 
\nonumber \\ && \qquad
F \,\, C_F\Big[4z-2(1-\hat{\xi})
+\varepsilon
[2z-(1-\hat{\xi})][1+\ln z+\ln(1-z)]
\Big]
\nonumber \\ && \qquad
+ F^2 \,\, \Big[\frac{1}{\varepsilon}\Big\{
C_F^2[ -16
          - 8z
          +16z(-2\ln(1 - z)+\ln z)
]
\nonumber \\ && \qquad
+C_AC_F[
\frac{20}{3}
          - \frac{88}{3}z]
+n_fC_FT_f[- \frac{16}{3}
          + \frac{32}{3}z]
\Big\}
\nonumber \\ && \qquad
+C_F^2[
- \frac{32}{3}- \frac{16}{3}z
          +\zeta(2)(-8+16z- \frac{32}{3}z^2)
          - 44z\ln(1 - z)
\nonumber \\ && \qquad
          - 24z\ln^2(1-z)
          -\frac{4}{3}\ln z(10-17z)
          + 8(1-4z)\ln z\ln(1 - z)
\nonumber \\ && \qquad
          +\frac{16}{3}(\frac{1}{z}-3z-2z^2)[\ln z\ln(1 + z)+{\rm Li}_2( - z)]
+ \ln^2 z(12z+\frac{16}{3}z^2)
\nonumber \\ && \qquad
          + 16{\rm Li}_2(1 - z)(1-3z)
          +8z(1-z)\{-2\ln(1 - z)\zeta(2)+2\ln z\zeta(2)
\nonumber \\ && \qquad
              -\ln^2 z\ln(1 - z)
-4\ln z{\rm Li}_2(1-z)
              -2\ln(1 - z){\rm Li}_2(1 - z)
\nonumber \\ && \qquad
 +6\zeta(3)+6{\rm Li}_3(1 - z)-6{\rm S}_{12}(1 - z)\}
]
\nonumber \\ && \qquad
+C_AC_F[
          -\frac{26}{9}+\frac{232}{9}z
          +4\zeta(2)(1-2z+\frac{4}{3}z^2) + \ln z(6-28z)
\nonumber \\ && \qquad
  -\frac{16}{3}\ln(1-z)(1+4z)
          - 4[\ln z\ln(1 - z)+2{\rm Li}_2(1 - z)](1-2z)
\nonumber \\ && \qquad
- \frac{8}{3}z^2\ln^2 z
          -\frac{8}{3}(\frac{1}{z}-3z-2z^2)[\ln z\ln(1+z)
            +{\rm Li}_2(-z)]
\nonumber \\ && \qquad
          + 8z(1-z)\{\frac{1}{2}\ln^2 z\ln(1-z)+\zeta(2)[\ln(1-z)
               -\ln z]+2\ln z{\rm Li}_2(1 - z)
\nonumber \\ && \qquad
  -3\zeta(3)
+\ln(1-z){\rm Li}_2(1 - z)
-3{\rm Li}_3(1 - z)+3{\rm S}_{12}(1 - z)\}
]
\nonumber \\ && \qquad
+n_fC_FT_f[-\frac{8}{9}(1+4z)
          - \frac{8}{3}(1-2z)\{\ln z+2\ln(1-z)\}
]
\nonumber \\ && \qquad
-\frac{16}{3}(-1)^n(C_F^2-\frac{1}{2}C_AC_F)\Big\{
1-z
          +z^2[-2\zeta(2)+\ln^2 z]
\nonumber \\ && \qquad
- \ln z(1-2z)
          +(\frac{1}{z}-3z-2z^2)
           [\ln z\ln(1+z)+{\rm Li}_2( - z)]
\Big\}
\Big]\,.
\end{eqnarray}
The purely singlet (PS) OME is split in the same way as the non-singlet
one in (\ref{eqn:2.7}). For the physical part we 
have (see Eq. (\ref{eqn:2.16}))
%(A5)
\begin{eqnarray}
\label{eqn:A5}
&&\hat A_{qq}^{\rm PS, PHYS}
\Big(z, \frac{-p^2}{\mu^2},\frac{1}{\varepsilon}\Big)
= 
\nonumber \\ && \qquad
F^2 \,\,n_fC_FT_f\Big[\frac{16}{\varepsilon^2}\Big\{
           1
          + \frac{4}{3z}
          - z
          - \frac{4}{3}z^2
          + 2\ln z(1+z)
          \Big\}
\nonumber \\ && \qquad
+\frac{8}{\varepsilon}\Big\{
           \frac{2}{3}(16
          + \frac{5}{z}
          - 13z
          - 8z^2)
          + 2\ln(1 - z)(1
          + \frac{4}{3z}
          - z
          - \frac{4}{3}z^2)
\nonumber \\ && \qquad
          + \ln z(13
          + \frac{8}{3z}
          + 13z)
          + 4(1+z)[\ln z\ln(1 - z)
          + \frac{3}{4}\ln^2 z
          + {\rm Li}_2(1 - z)]
          \Big\}
\nonumber \\ && \qquad
  +\frac{316}{3}
          + \frac{20}{z}
          - \frac{256}{3}z
          - 40z^2
          + 4\{\zeta(2)+2\ln(1-z)\}(1+\frac{4}{3z}-z- \frac{4}{3}z^2)
\nonumber \\ && \qquad
          + \frac{16}{3}\ln(1 - z)(16
           + \frac{5}{z}
           - 13z
           - 8z^2)
   + 2\ln^2 z(37
           + \frac{16}{3z}
           + 41z
           + \frac{8}{3}z^2)
\nonumber \\ && \qquad
          + 8\ln z\ln(1 - z)(13
           + \frac{8}{3z}
           + 13z)
  + \frac{8}{3}\ln z(67
           + \frac{10}{z}
           + 50z)
\nonumber \\ && \qquad
          + 16{\rm Li}_2(1 - z)(6
            + \frac{2}{3z}
             + 7z+ \frac{2}{3}z^2)
          +16(1+z)[\frac{1}{2}\ln z\zeta(2)
\nonumber \\ && \qquad
+2\ln(1 - z){\rm Li}_2(1 - z)
 +\ln z\ln^2(1-z
+\frac{3}{2}\ln^2 z\ln(1 - z)
\nonumber \\ && \qquad
           +\frac{7}{12}\ln^3 z+2\ln z{\rm Li}_2(1 - z)
           - 2{\rm Li}_3(1 - z)
           + {\rm S}_{12}(1 - z)]
\Big]\,,
\end{eqnarray}
and the unphysical part (see Eq. (\ref{eqn:2.17})) becomes equal to
%(A6)
\begin{eqnarray}
\label{eqn:A6}
&&\hat A_{qq}^{\rm PS, EOM}
\Big(z, \frac{-p^2}{\mu^2},\frac{1}{\varepsilon}\Big)
= 
\nonumber \\ && \qquad
F^2 \,\, n_fC_FT_f\Big[\frac{32}{\varepsilon}\Big\{
          - 2
          + z
          + z^2
          - \ln z(1+2z)
          \Big\}
\nonumber \\ && \qquad
          - 72
          + 32z
          + 40z^2
          +32\ln(1 - z)(-2+z+z^2)
          -16\ln z(6+7z)
\nonumber \\ && \qquad
          - 32(1+2z)[\ln z\ln(1 - z)+\frac{3}{4}\ln^2 z+{\rm Li}_2(1 - z)]
\Big]\,.
\end{eqnarray}
The next OME is split into three pieces according to Eq. (\ref{eqn:2.18}).
The physical part is given by (see Eq. (\ref{eqn:2.27}))
%(A.7)
\begin{eqnarray}
\label{eqn:A.7}
&&\hat A_{qg}^{\rm PHYS}   
\big(z, \frac{-p^2}{\mu^2},\frac{1}{\varepsilon}\Big)
=
\nonumber \\ && \qquad
F\,\, n_f T_f\Big[\frac{8}{\varepsilon}
\Big\{1-2z+2z^2\Big\}
+4+4(1-2z+2z^2)[\ln z+\ln(1-z)]
\nonumber \\ && \qquad
+\varepsilon\Big\{2[\ln z+\ln(1-z)]+(1-2z+2z^2)[\zeta(2)+\{\ln z
+\ln(1-z)\}^2]\Big\}\Big]
\nonumber \\ && \qquad
+F^2 \,\, n_f\Big[\frac{1}{\varepsilon^2}\Big\{
C_AT_f[\frac{16}{3}(14+\frac{4}{z}+2z
          -9z^2)+ 32\ln z  (
           1
          + 4z
          )
\nonumber \\ && \qquad
+ 32\ln(1 - z)  (
           1
          - 2z
          + 2z^2
          )
]-n_fT_f^2\frac{64}{3}[1-2z+2z^2]
\nonumber \\ && \qquad
-8C_FT_f[
1-4z
 -4\ln(1 - z)  (
           1
          - 2z
          + 2z^2
          )
       +2\ln z  (
           1
          - 2z
          + 4z^2
          )
]\Big\}
\nonumber \\ && \qquad
+\frac{1}{\varepsilon}\Big\{
C_AT_f[\frac{8}{9}(26
          +\frac{30}{z}
          -79z
          -20z^2)
+24\ln^2(1-z)(1-2z+2z^2)
\nonumber \\ && \qquad
 + 32\zeta(2)(
           1
          - z
          + 2z^2
          )
       + \frac{8}{3}\ln(1 - z)  (
           17
          + \frac{8}{z}
          + 38z
          - 52z^2
          )+32(1+4z)
\nonumber \\ && \qquad
       \times \ln z\ln(1 - z)
       + 8\ln^2 z  (
           3
          + 10z
          )
       + \frac{16}{3}\ln z  (
           25
          + \frac{4}{z}
          + 13z
          + 14z^2
          )
\nonumber \\ && \qquad
       + 64z(3-z){\rm Li}_2(1 - z)
       + 16(
           1
          + 2z
          + 2z^2
          )[\ln z\ln(1+z)+{\rm Li}_2( - z)]
]
\nonumber \\ && \qquad
+C_FT_f[
 - 32
          + 36z
 + 8(
           1
          - 2z
          + 2z^2
          )[-2\zeta(2)+3\ln^2(1-z)]
\nonumber \\ && \qquad
       -4 \ln z  (
          5
          + 4z
          + 8z^2
          )
 + 8\ln(1 - z)  (
           3
          + 4z^2
          )
  -12\ln^2 z  (
          1
          - 2z
          + 4z^2
          )
\nonumber \\ && \qquad
 + 32\ln z\ln(1 - z)  (
           1
          - 2z
          + z^2
          )
       + 16{\rm Li}_2(1 - z)  (
           3
          - 6z
          + 4z^2
          )
]
\nonumber \\ && \qquad
+n_fT_f^2\frac{32}{3}[\frac{1}{3}(2-10z+10z^2)
-(1-2z+2z^2)\{\ln z+\ln(1-z)\}] \Big\}
\nonumber \\ && \qquad
+C_AT_f[
\frac{2900}{27}
          + \frac{20}{z}
          - \frac{136}{27}z
          - \frac{2492}{27}z^2
       +\zeta(2)(12+\frac{16}{3z}+16z-\frac{172}{3}z^2
\nonumber \\ && \qquad
      +16z^3) +8\zeta(3)(1-z-4z^2+6z^3)
       + \frac{28}{3}\ln^3(1-z)  (
           1
          -2z
          +2z^2
          )
\nonumber \\ && \qquad
       + \frac{2}{3}\ln^2(1-z)  (
           23
          +\frac{16}{z}
          + 110z
          -138z^2
          )
       + 8\ln(1 - z){\rm Li}_2(1 - z)  (
           4
\nonumber \\ && \qquad
          +23z+ 2z^2
          - 2z^3
          )
       + \frac{4}{9}\ln(1 - z)  (
           68
          +\frac{60}{z}
          -283z
          +94z^2
          )
\nonumber \\ && \qquad
       +8\ln(1-z)\zeta(2)(3-7z+8z^2-2z^3)
       + 8\ln z\ln^2(1-z)  (
           3
          + 10z
          + 2z^2
          )
\nonumber \\ && \qquad
       + \frac{4}{3}\ln z\ln(1 - z)  (
           83
          + \frac{16}{z}
          + 86z
          + 22z^2
          )
       + 4\ln^2 z\ln(1 - z)  (
           5
          + 21z
\nonumber \\ && \qquad
      -2z^3) + \frac{4}{3}\ln^3 z  (
           7
          +22z
          )
       + \frac{4}{3}\ln^2 z  (
           61
          +\frac{8}{z}
          +49z
          +51z^2
          -6z^3
          )
\nonumber \\ && \qquad
       + 16\ln z{\rm Li}_2(1 - z)  (
           1
          + 7z
          - 2z^3
          )
       + \frac{8}{9}\ln z  (
          157
          +\frac{30}{z}
          +127z
          +59z^2
          )
\nonumber \\ && \qquad
       +8\ln z\zeta(2)(2+3z+2z^3)
       + 16{\rm Li}_2(1 - z)  (
           5
          + \frac{2}{3z}
          + 4z
          + \frac{37}{6}z^2
          )
\nonumber \\ && \qquad
       -8 (
           1
          + 4z
          + z^2
          - 2z^3
          )[\ln z\ln(1+z)+{\rm Li}_2( - z)]
\nonumber \\ && \qquad
       - 8{\rm Li}_3(1 - z)  (
           8
          + 13z
          + 14z^2
          - 6z^3
          )
       + 8{\rm S}_{12}(1 - z)  (
           5
          - 5z
          + 12z^2
\nonumber \\ && \qquad
-6z^3)       +8(
          1
          + 2z
          + 2z^2
          )\{2\ln(1+z){\rm Li}_2(1-z)+4\ln(1 + z){\rm Li}_2(-z)
\nonumber \\ && \qquad
            +2\ln(1 + z)\zeta(2)
+2\ln z\ln(1 - z)\ln(1 + z)+2\ln z\ln^2(1+z)
\nonumber \\ && \qquad
            +\frac{1}{2}\ln^2 z\ln(1 + z)
+\ln z{\rm Li}_2( - z)
            -{\rm Li}_3( - z)+6{\rm S}_{12}( - z)-{\rm S}_{12}(z^2)\}
]
\nonumber \\ && \qquad
+C_FT_f[
 - 34
          + 58z
       -2\zeta(2)(9-12z+8z^2)
\nonumber \\ && \qquad
       +4(
           1
          - 2z
          +2z^2
          )[-10\zeta(3)+\frac{7}{3}\ln^3(1-z)-2\ln(1-z)\zeta(2)]
\nonumber \\ && \qquad
       + 4\ln^2(1-z)  (
           5
          - 2z
          + 6z^2
          )
       + 8\ln(1 - z){\rm Li}_2(1 - z)  (
           5
          - 10z
          + 6z^2
          )
\nonumber \\ && \qquad
       -16z(1-3z)\ln(1 - z) 
       + 4\ln z\ln^2(1-z)  (
           7
          - 14z
          + 10z^2
          )
\nonumber \\ && \qquad
       + 4\ln z\ln(1 - z)  (
           7
          - 16z
          + 4z^2
          )
       + 4\ln^2 z\ln(1 - z)  (
           5
          - 10z
          + 4z^2
          )
\nonumber \\ && \qquad
       -\frac{14}{3}\ln^3 z  (
           1
          - 2z
          + 4z^2
          )
       - \ln^2 z  (
           13
          + 20z
          + 24z^2
          )
\nonumber \\ && \qquad
       + 32\ln z{\rm Li}_2(1 - z)  (
           2
          - 4z
          + 3z^2
          )
       - 2\ln z  (
           34
          - 15z
          + 24z^2
          )
\nonumber \\ && \qquad
       -4\ln z\zeta(2)(5-10z+12z^2)
       +8(
           3
          - 6z
          + 2z^2
          )[2{\rm Li}_2(1 - z) 
\nonumber \\ && \qquad
-{\rm Li}_3(1 - z)]
       + 16{\rm S}_{12}(1 - z)  (
           3
          - 6z
          + 5z^2
          )
]
\nonumber \\ && \qquad
-n_fT_f^2\frac{8}{3}[
\frac{2}{9}(13-56z+56z^2)
-\frac{4}{3}(1-5z+5z^2)[\ln z +\ln(1-z)]
\nonumber \\ && \qquad
+(1-2z+2z)\{\ln z+\ln(1-z)\}^2]
\Big\}\Big]\,.
\end{eqnarray}
The unphysical part (EOM) (see Eq. (\ref{eqn:2.28})) becomes equal to
%(A8)
\begin{eqnarray}
\label{eqn:A8}
&&\hat A_{qg}^{\rm EOM}
\Big(z, \frac{-p^2}{\mu^2},\frac{1}{\varepsilon}\Big)
=
\nonumber \\ && \qquad
F\,\, n_fT_f\Big[-16z(1-z)-8\varepsilon z(1-z)[\ln z+\ln(1-z)]\Big]
\nonumber \\ && \qquad
+ F^2\,\, n_f\Big[\frac{1}{\varepsilon}\Big\{
C_AT_f[
48
          + \frac{128}{3}z
          - \frac{272}{3}z^2
 + 16\ln z  (
           1
          + 8z
          + 8z^2)
]
\nonumber \\ && \qquad
+C_FT_f[
-32+32z    - 128z(1-z)\ln(1 - z)
       + 64z(1-2z)\ln z]
\nonumber \\ && \qquad
+n_fT_f^2\frac{128}{3}z(1-z)
\Big\}
\nonumber \\ && \qquad
+C_AT_f[
 68
          - \frac{296}{9}z
          -\frac{316}{9}z^2
       +32\zeta(2)(-z+z^2+z^3)
\nonumber \\ && \qquad
       +16\zeta(3)(z-11z^2+6z^3)
       + 16\ln(1 - z)  (
           3
          + \frac{25}{3}z
          - \frac{34}{3}z^2
          )
\nonumber \\ && \qquad
       +16\ln(1-z)\zeta(2)(z+z^2-2z^3)
       + 16\ln z\ln(1 - z)  (
           1
          + 10z
          + 6z^2
          )
\nonumber \\ && \qquad
       + 4\ln^2 z  (
           3
          + 24z
          + 24z^2
          - 4z^3       
)
     -32z(1-z)\ln z{\rm Li}_2( - z)
\nonumber \\ && \qquad
       + 8\ln z  (
           7
          + \frac{50}{3}z
          + \frac{34}{3}z^2
          )
       + 16{\rm Li}_2(1 - z)  (
           1
          + 12z
          + 4z^2
          )
\nonumber \\ && \qquad
       + 32(
          - 1
          + 2z^2
          + z^3
          )[\ln z\ln(1+z)+{\rm Li}_2( - z)]
       + 16(
           z
          - 3z^2
          + 2z^3
          )
\nonumber \\ && \qquad
\times[-\ln(1-z){\rm Li}_2(1-z)-\frac{1}{2}\ln^2 z\ln(1 - z)
             -2\ln z{\rm Li}_2(1 - z)
\nonumber \\ && \qquad
             +\ln z\zeta(2)+3{\rm Li}_3(1 - z)]
       -16 {\rm S}_{12}(1 - z)  (
           5z
          - 11z^2
          + 6z^3
          )
\nonumber \\ && \qquad
+ 32z(1-3z){\rm Li}_3( - z)
   + 32z(1+z)\{-\frac{1}{2}\ln^2 z\ln(1 + z)
\nonumber \\ && \qquad
+\ln z\ln^2(1+z)+\ln(1+z)\zeta(2)+2\ln(1+z){\rm Li}_2(-z)
+2{\rm S}_{12}(-z)\} ]
\nonumber \\ && \qquad
+C_FT_f[
 - 16(1-z)
       +32z(1-z)[2\zeta(2)-3\ln^2(1-z)]
\nonumber \\ && \qquad
       -32 \ln(1 - z)  (
           1
          + 3z
          - 4z^2
          )
       -64z(2-z) \ln z\ln(1 - z)
\nonumber \\ && \qquad
       +48z(1-2z) \ln^2 z
       -16 \ln z  (
           1
          - z
          +8z^2
          )
       -64z(3-2z) {\rm Li}_2(1 - z)
]
\nonumber \\ && \qquad
+n_fT_f^2\frac{64}{3}z(1-z)[-\frac{5}{3}+\ln z+\ln(1-z)]
\Big]\,.
\end{eqnarray}
%%%%%%%%%%%%%%%%%%%%%%%%%%%%%%
The part which is due to the contribution of the non-gauge invariant (NGI) 
operator $O_A$ (see Eq. (\ref{eqn:2.29})) is equal to
%(A9)
\begin{eqnarray}
\label{eqn:A9}
&&\hat A_{qg}^{\rm NGI}
\Big(z, \frac{-p^2}{\mu^2},\frac{1}{\varepsilon}\Big)
= 
F^2\,\, n_fC_AT_f\Big[-\frac{16}{\varepsilon^2}\Big\{
\frac{2}{3z}
          -z
          + \frac{1}{3}z^2
+\ln z\Big\}
\nonumber \\ && \qquad
+\frac{1}{\varepsilon}\Big\{
-\frac{8}{3}(7
          +\frac{5}{z}
          - 13z
          + z^2)
 - \frac{16}{3}\ln(1 - z)  (
           \frac{2}{z}
          - 3z
          + z^2
          )
\nonumber \\ && \qquad
       -16 \ln z\ln(1 - z)  
       -12 \ln^2 z  
       - 8\ln z  (
           5
          + \frac{4}{3z}
          + z
          )
 -16 {\rm Li}_2(1 - z)\Big\}
\nonumber \\ && \qquad
- \frac{104}{3}
          - \frac{10}{z}
          + \frac{122}{3}z
          + 4z^2
       -\frac{4}{3}(\frac{2}{z}-3z+z^2)[\zeta(2)+2\ln^2(1-z)]
\nonumber \\ && \qquad
       -16\ln(1 - z){\rm Li}_2(1 - z)
       -\frac{8}{3}\ln(1 - z)  (
           7
          + \frac{5}{z}
          - 13z
          + z^2
          )
\nonumber \\ && \qquad
       -8 \ln z\ln^2(1-z)
       -8 \ln z\ln(1 - z)  (
           5
          + \frac{4}{3z}
          + z
          )
       -12 \ln^2 z\ln(1 - z)
\nonumber \\ && \qquad
       - \frac{14}{3} \ln^3 z
       -2 \ln^2 z  (
           15
          + \frac{8}{3z}
          + 5z
          - \frac{2}{3}z^2
          )
       -16 \ln z{\rm Li}_2(1 - z)
\nonumber \\ && \qquad
       - \frac{8}{3}\ln z  (
          23
          + \frac{5}{z}
          + 10z
          )
       -4 {\rm Li}_2(1 - z)  (
           10
          + \frac{4}{3z}
          + 4z
          - \frac{2}{3}z^2
          )
\nonumber \\ && \qquad
 -4\ln z\zeta(2)
       + 16{\rm Li}_3(1 - z)
       -8 {\rm S}_{12}(1 - z)
\Big]\,.
\end{eqnarray}
%%%%%%%%%%%%%%%%%%%%%%%%%%%%%%%%%%%%%%%
The next OME is split into physical and unphysical parts according
to Eq. (\ref{eqn:2.30}). The former (see Eq. (\ref{eqn:2.31})) is
%(A10)
\begin{eqnarray}
\label{eqn:A10}
&&\hat A_{gq}^{\rm PHYS}
\Big(z, \frac{-p^2}{\mu^2},\frac{1}{\varepsilon}\Big)
= 
\nonumber \\ && \qquad
F\,\, C_F\Big[\frac{4}{\varepsilon}
\Big\{z-2+\frac{2}{z}\Big\}
-6+\frac{4}{z}+6z
-(4-\frac{4}{z}-2z)[\ln z+\ln(1-z)]
\nonumber \\ && \qquad
+\varepsilon\Big\{-1+2z
+(-3+\frac{2}{z}+3z)[\ln z+\ln(1-z)]
\nonumber \\ && \qquad
+(-1+\frac{1}{z}+\frac{z}{2})[\zeta(2)+\{\ln z+\ln(1-z)\}^2]\Big\}\Big]
\nonumber \\ && \qquad
+ F^2 \,\,\Big[\frac{1}{\varepsilon^2}\Big\{
C_F^2[
 16
          - 4z
-16 \ln(1 - z)  (
          2
          - \frac{2}{z}
          - z
          )
       + 8\ln z  (
           2
          - z
          )
]
\nonumber \\ && \qquad
+C_AC_F[
\frac{16}{3}
          - \frac{24}{z}
          + \frac{112}{3}z
          + \frac{32}{3}z^2
 + 16\ln(1 - z)  (
          - 2
          + \frac{2}{z}
          + z
          )
\nonumber \\ && \qquad
       -32 \ln z  (
           1
          + \frac{1}{z}
          + z
          )
]
-n_fC_FT_f\frac{32}{3}[
-2
          + \frac{2}{z}
          + z
]
\Big\}
\nonumber \\ && \qquad
+\frac{1}{\varepsilon}\Big\{
C_F^2[
34-38z
      +4(
          - 2
          + \frac{2}{z}
          + z
          )\{4\zeta(2)+3\ln^2(1-z)\}
\nonumber \\ && \qquad
       + 4\ln(1 - z)  (
           4
          + z
          )
  + 2(
           2
          - z
          )\{4\ln z\ln(1 - z)+3\ln^2 z\}
\nonumber \\ && \qquad
       + 2\ln z  (
           24
          + 5z
          )
   - 8{\rm Li}_2(1 - z)  (
          - 6
          + \frac{4}{z}
          +3z
          )
]
\nonumber \\ && \qquad
+C_AC_F[
- \frac{604}{9}
          - \frac{212}{9z}
          + \frac{668}{9}z
          + \frac{80}{3}z^2
       -16\zeta(2)  (
          - 1
          +\frac{2}{z}
          + z
          )
\nonumber \\ && \qquad
       + 12\ln^2(1-z)  (
          - 2
          + \frac{2}{z}
          + z
          )
       - \frac{4}{3}\ln(1 - z)  (
           10
          +\frac{16}{z}
          -47z
          -8z^2
          )
\nonumber \\ && \qquad
       -8 \ln z\ln(1 - z)  (
           10
          - \frac{2}{z}
          + z
          )
       - \ln z  (
           \frac{296}{3}
          + \frac{24}{z}
          + \frac{212}{3}z
          )
\nonumber \\ && \qquad
    -4 \ln^2 z  (
           6
          + \frac{4}{z}
          + 5z
          )
       -32 {\rm Li}_2(1 - z)  (
          3
          - \frac{1}{z}
          )
\nonumber \\ && \qquad
       -8(
          2
          +\frac{2}{z}
          +z
          )\{\ln z\ln(1+z)+{\rm Li}_2( - z)\}
]
\nonumber \\ && \qquad
+n_fC_FT_f[
\frac{32}{9}(4
          - \frac{1}{z}
          - 5z)
   +\frac{16}{3}(
          2
          -\frac{2}{z}
          -z
          )\{\ln(1-z)+2\ln z\}]\Big\}
\nonumber \\ && \qquad
+C_F^2[
38
          - 29z
       + \zeta(2)  (
          - 16
          + \frac{4}{z}
          + 17z
          )
       + 16\zeta(3)  (
           1
          - \frac{1}{z}
          + z
          )
\nonumber \\ && \qquad
       + 4\ln(1 - z)\zeta(2)  (
          - 10
          + \frac{10}{z}
          + 3z
          )
       +\frac{14}{3}\ln^3(1-z)  (
          - 2
          + \frac{2}{z}
          + z
          )
\nonumber \\ && \qquad
 + 4\ln^2(1-z)  (
           2
          + z
          )
       -8\ln(1 - z){\rm Li}_2(1 - z)  (
           -4
          +\frac{2}{z}
          +3z
          )
\nonumber \\ && \qquad
 +4\ln(1 - z)  (
          13
          -\frac{2}{z}
          -11z
          )
       +4 \ln z\ln(1 - z)  (
           17
          -\frac{1}{z}
          )
\nonumber \\ && \qquad
       + 2\ln^2 z\ln(1 - z)  (
           6
          - 5z
          )
       + (
           2
          - z
          )\{4 \ln z\ln^2(1-z)+\frac{7}{3}\ln^3 z\}
\nonumber \\ && \qquad
       + \ln^2 z  (
           32
          + \frac{17}{2}z
          )
       + 8\ln z{\rm Li}_2(1 - z)  (
           2
          - 3z
          )
       + 33(1+z)\ln z
\nonumber \\ && \qquad
       + 2\ln z\zeta(2)  (
           2
          + 3z
          )
       -8{\rm Li}_2(1 - z)  (
           -10
          +\frac{1}{z}
          +2z
          )
\nonumber \\ && \qquad
       -8{\rm Li}_3(1 - z)  (
          \frac{2}{z}
          -3z
          )
       -4 {\rm S}_{12}(1 - z)  (
          6
          -\frac{8}{z}
          +3z
          )
]
\nonumber \\ && \qquad
+C_AC_F[
\frac{1}{27}(319
          - \frac{2539}{z}
          + 1678z
          + 828z^2)
       -4 \zeta(3)  (
           8
          -\frac{1}{z}
          -11z
          +6z^2
          )
\nonumber \\ && \qquad
  + \frac{2}{3}\zeta(2)  (
            20
          - \frac{15}{z}
          - 16z
          + 4z^2
          )
       +4 \ln(1 - z)\zeta(2)  (
           10
          - \frac{7}{z}
          - 7z
          + 2z^2
          )
\nonumber \\ && \qquad
       -\frac{14}{3} \ln^3(1-z)  (
          2
          - \frac{2}{z}
          - z
          )
       - \frac{1}{3}\ln^2(1-z)  (
          34
          + \frac{30}{z}
          -113z
          -16z^2
          )
\nonumber \\ && \qquad
       -4 \ln(1 - z){\rm Li}_2(1 - z)  (
          22
          -\frac{1}{z}
          +7z
          -2z^2
          )
\nonumber \\ && \qquad
       - \frac{2}{9}\ln(1 - z)  (
           454
          - \frac{22}{z}
          - 425z
          - 120z^2
          )
       -4 \ln z\zeta(2)  (
           4
          + \frac{3}{z}
          - z
          + 2z^2
          )
\nonumber \\ && \qquad
       -2 \ln z\ln^2(1-z)  (
           30
           -\frac{6}{z}
           +z
          )
       - \frac{4}{3}\ln z\ln(1 - z)  (
           97
          + \frac{10}{z}
          + 25z
          )
\nonumber \\ && \qquad
       -2\ln^2 z\ln(1 - z)  (
           22
          -\frac{5}{z}
          +7z
          -2z^2
          )
       -\frac{2}{3} \ln^3 z  (
          14
          +\frac{8}{z}
          +11z
          )
\nonumber \\ && \qquad
       - \frac{1}{3}\ln^2 z  (
           226
          + \frac{42}{z}
          + 175z
          + 8z^2
          )
       -8 \ln z{\rm Li}_2(1 - z)  (
           8
          -\frac{5}{z}
          +3z
          -2z^2
          )
\nonumber \\ && \qquad
       -\frac{1}{9} \ln z  (
          1546
          +\frac{680}{z}
          +856z
          )
       -4 {\rm Li}_2(1 - z)  (
           36
          - \frac{5}{3z}
          + 10z
          + \frac{4}{3}z^2
          )
\nonumber \\ && \qquad
       + 4(
           1
          - \frac{1}{z^2}
          + \frac{1}{z}
          - 2z
          )\{\ln z\ln(1+z)+{\rm Li}_2( - z)\}
\nonumber \\ && \qquad
       + 4{\rm Li}_3(1 - z)  (
           10
          + \frac{5}{z}
          +17z
          - 6z^2
          )
       -4{\rm S}_{12}(1 - z)  (
           4
          - \frac{9}{z}
          + 8z
          - 6z^2
          )
\nonumber \\ && \qquad
       + 4(
           2
          + \frac{2}{z}
          + z
          )\{-2\ln(1 + z)\zeta(2)
            -2\ln(1 + z){\rm Li}_2(1 - z)
\nonumber \\ && \qquad
-4\ln(1+z){\rm Li}_2(-z)
            -2\ln z\ln(1 - z)\ln(1 + z)-2\ln z\ln^2(1+z)
\nonumber \\ && \qquad
            -\frac{1}{2}\ln^2 z\ln(1 + z)
-\ln z{\rm Li}_2( - z) 
            +{\rm Li}_3( - z)-6{\rm S}_{12}( - z)+{\rm S}_{12}(z^2)\}
]
\nonumber \\ && \qquad
+n_fC_FT_f[
\frac{8}{27}(38
          - \frac{32}{z}
          - 25z)
       +\frac{4}{3}(
          2
          -\frac{2}{z}
          -z
          )[\{2\ln z+\ln(1-z)\}^2
\nonumber \\ && \qquad
+2\zeta(2)]
       +\frac{16}{9}(
          4
          -\frac{1}{z}
          -5z
          )\{\ln(1-z)+2\ln z\}]
     \Big ]\,.
\end{eqnarray}
%%%%%%%%%%%%%%%%%%%%%%%%%%%%%%%%%%%%%%%%%%
The unphysical part is given by (see Eq. (\ref{eqn:2.32}))
%(A11)
\begin{eqnarray}
\label{eqn:A11}
&&\hat A_{gq}^{\rm EOM}
\Big(z, \frac{-p^2}{\mu^2},\frac{1}{\varepsilon}\Big)
= 
\nonumber \\ && \qquad
F\,\, C_F\Big[4(1-z)+2\varepsilon(1-z)[1+\ln z+\ln(1-z)]\Big]
\nonumber \\ && \qquad
+ F^2 \,\, \Big [\frac{1}{\varepsilon}\Big\{
16C_F^2[
-2
          + \frac{1}{z}
          + z
-z\ln z
]-n_fC_FT_f\frac{64}{3}(1-z)
\nonumber \\ && \qquad
+16C_AC_F[
\frac{14}{3}(1-z)
          +\frac{1}{z}-z^2
 + 2(1-z)\ln(1 - z)
       + 2(1+2z)\ln z
]
\Big\}
\nonumber \\ && \qquad
+ C_F^2[
-8(1-\frac{1}{z})
+ 16(1-z)\zeta(2)
       -4\ln(1 - z)  (
           9
         - \frac{4}{z}
          - 5z
          ) -12z \ln^2 z
\nonumber \\ && \qquad
       -16 \ln z\ln(1 - z)
       -8 \ln z  (
          2
          -\frac{1}{z}
          +2z
          )
       -16(2-z) {\rm Li}_2(1 - z)
]
\nonumber \\ && \qquad
+C_AC_F[
\frac{400}{9}
          - \frac{28}{3z}
          - \frac{64}{9}z
          - 28z^2
       + 24(1-z)\ln^2(1-z)
\nonumber \\ && \qquad
       + 16\ln(1 - z)  (
           \frac{16}{3}
          + \frac{1}{z}
          - \frac{16}{3}z
          - z^2
          )
       + 24(3+z)\ln z\ln(1 - z)
\nonumber \\ && \qquad
         +8\ln^2 z  (
           3
          + \frac{16}{3}z
          )
       + 2\ln z  (
           55
          + \frac{2}{3z}
          + \frac{80}{3}z
          )
       + 16(5+z){\rm Li}_2(1 - z)
\nonumber \\ && \qquad
 -8 \zeta(2)  (
          1
          - \frac{7}{3}z
          )
       + 4(
           3
          - \frac{1}{3z^2}
          + \frac{8}{3}z
          )\{\ln z\ln(1+z)+{\rm Li}_2( - z)\}
\nonumber \\ && \qquad
       -4(
           1
          - 3z
          + 2z^2
          )\{-\ln z\zeta(2)+\ln(1 - z){\rm Li}_2(1 - z)
\nonumber \\ && \qquad
        +\frac{1}{2}\ln^2 z\ln(1 - z)
 + \ln(1-z)\zeta(2)-3\zeta(3)
            +2\ln z{\rm Li}_2(1 - z) 
\nonumber \\ && \qquad
 -3{\rm Li}_3(1 - z)
+3{\rm S}_{12}(1 - z)\}
]
\nonumber \\ && \qquad
-n_fC_FT_f\frac{32}{3}(1-z)[
\frac{1}{3}+\ln(1 - z)+2\ln z]
\Big\}
\Big]\,.
\end{eqnarray}
%%%%%%%%%%%%%%%%%%%%%%%%%%%%%%%%%%%%%%%%%%%%%%%
The gluonic OME can be split into three parts according to Eq. 
(\ref{eqn:2.33}). The physical part is equal to (see Eq. (\ref{eqn:2.34}))
%(A12)
\begin{eqnarray}
\label{eqn:A12}
&&\hat A_{gg}^{\rm PHYS}
\Big(z, \frac{-p^2}{\mu^2},\frac{1}{\varepsilon}\Big)
= \delta(1-z)
\nonumber \\ && \qquad
+ F \,\, \Big[\frac{1}{\varepsilon}
\Big\{
C_A[-16+\frac{8}{z}+8z-8z^2+\frac{8}{1-z}
\nonumber \\ && \qquad
+\frac{22}{3}\delta(1-z)]-\frac{8}{3}n_fT_f\delta(1-z)\Big\}
+C_A[-6+\frac{4}{z}+4z-4z^2
\nonumber \\ && \qquad
-4(2-\frac{1}{z}-z+z^2+\frac{4}{1-z})[\ln z+\ln(1-z)]
+\frac{1-\hat{\xi}}{1-z}
\nonumber \\ && \qquad
+\delta(1-z)\{-\frac{67}{9}+4\zeta(2)-(1-\hat{\xi})
+\frac{1}{4}(1-\hat{\xi})^2\}]
+n_fT_f\frac{20}{9}\delta(1-z)
\nonumber \\ && \qquad
+\varepsilon\Big\{C_A[(-2+\frac{1}{z}+z-z^2+\frac{1}{1-z})[\zeta(2)
+\{\ln z+\ln(1-z)\}^2]
\nonumber \\ && \qquad
-(3-\frac{2}{z}-2z+2z^2)[\ln(1-z)+\ln z]
-\frac{1}{2}\frac{1-\hat{\xi}}{1-z}\{1
-\ln z-\ln(1-z)\}
\nonumber \\ && \qquad
+\delta(1-z)(\frac{202}{27}-\frac{11}{12}\zeta(2)-\frac{14}{3}\zeta(3)
+\frac{1}{4}\zeta(2)(1-\hat{\xi})-\frac{1}{4}(1-\hat{\xi})^2)]
\nonumber \\ && \qquad
-n_fT_f\delta(1-z)(\frac{56}{27}-\frac{1}{3}\zeta(2))\Big]
\nonumber \\ && \qquad
+ F^2 \,\, \Big [\frac{1}{\varepsilon^2}\Big\{C_A^2[
 - 8(10+z)
          - \frac{88}{3}(\frac{1}{z}-z^2)
          + \frac{88}{1-z}
\nonumber \\ && \qquad
 + \delta(1-z) \{
           \frac{484}{9}
          - 32\zeta(2)
          \}
   - 64\ln(1 - z)  (
           2
          - \frac{1}{z}
          - z
          + z^2
\nonumber \\ && \qquad
          - \frac{1}{1-z}
          )
       - 32\ln z  (
           \frac{1}{z}
          + 3z
          - z^2
          + \frac{1}{1-z}
          )
]
\nonumber \\ && \qquad
+n_fC_AT_f[
  32(2
          - \frac{1}{z}
          - z
          + z^2
          - \frac{1}{1-z}) -\frac{352}{9} \delta(1-z)]
\nonumber \\ && \qquad
+16n_fC_FT_f[
 1-z
          + \frac{4}{3}(\frac{1}{z}-z^2)
 + 2\ln z  (
           1
          + z
          )
]
+n_f^2T_f^2\frac{64}{9}\delta(1-z)
\Big\}
\nonumber \\ && \qquad
+\frac{1}{\varepsilon}\Big\{
C_A^2[
\frac{2}{3}(
 53
          -\frac{46}{z}
          +55z
          +28z^2)
          -\frac{86}{1-z}
 -8\zeta(2)(4+2z^2
\nonumber \\ && \qquad
-\frac{1}{1+z}
-\frac{1}{1-z})
 + \delta(1-z)  \{
          - \frac{3326}{27}
          + \frac{176}{3}\zeta(2)
          + 20\zeta(3)
          \}
\nonumber \\ && \qquad
- 48\ln^2(1-z)  (
           2
          - \frac{1}{z}
          - z
          + z^2
          - \frac{1}{1-z}
          )
 -\frac{16}{3}\ln(1 - z)  (
          13+\frac{5}{z}
\nonumber \\ && \qquad
          +z
          -5z^2
          -\frac{11}{1-z}
          )
 - 16\ln z\ln(1 - z)  (
           6
          - \frac{1}{z}
          + 3z
          + z^2
          - \frac{1}{1-z}
          )
\nonumber \\ && \qquad
 - 4\ln^2 z  (
           \frac{4}{z}
          + 16z
          - 6z^2
          + \frac{1}{1+z}
          + \frac{5}{1-z}
          )
\nonumber \\ && \qquad
  -4 \ln z  (
          36
          +\frac{6}{z}
          +21z
          - \frac{44}{3}\frac{1}{1-z}
          )
 -64 (1+z){\rm Li}_2(1 - z)
\nonumber \\ && \qquad
 -16(
           2
          + \frac{1}{z}
          + z
          + z^2
          - \frac{1}{1+z}
          )\{\ln z\ln(1+z)+{\rm Li}_2( - z)\}
]
\nonumber \\ && \qquad
+n_fC_AT_f[
 - \frac{8}{3}(11-5z)
          + \frac{152}{9}(\frac{1}{z}-z^2)
          +\frac{24}{1-z}
\nonumber \\ && \qquad
          +\frac{64}{3} \ln(1 - z)  (
           2
          - \frac{1}{z}
          - z
          + z^2
          - \frac{1}{1-z}
          )
\nonumber \\ && \qquad
+ \frac{16}{3}\ln z  (
           9
          - \frac{4}{z}
          - 3z
          + 4z^2
          - \frac{4}{1-z}
          )
 + \delta(1-z)  \{
           \frac{2168}{27}
          - \frac{64}{3}\zeta(2)\}
]
\nonumber \\ && \qquad
+n_fC_FT_f[
 \frac{160}{3}(1-z)
          - \frac{176}{9}(\frac{1}{z}-z^2)
  +4\delta(1-z)
\nonumber \\ && \qquad
 + 16\ln(1 - z)  (
           1
          + \frac{4}{3z}
          - z
          - \frac{4}{3}z^2
          )
  + 8\ln z  (
           5
          + 3z
          - \frac{8}{3}z^2
          )
\nonumber \\ && \qquad
       + 8(1+z)\{4\ln z\ln(1 - z)+3\ln^2 z+4{\rm Li}_2(1 - z)\}
]
\nonumber \\ && \qquad
-n_f^2T_f^2\frac{320}{27}\delta(1-z)
\Big\}
\nonumber \\ && \qquad
+C_A^2[
- \frac{1672}{45}
          - \frac{13951}{135z}
          + \frac{187}{45}z
          + \frac{8686}{135}z^2
          + \frac{254}{3}\frac{1}{1-z}
\nonumber \\ && \qquad
          +\zeta(2)(\frac{1}{3}-\frac{40}{3z}
            -\frac{61}{3}z+26z^2-\frac{124}{15}z^3+\frac{19}{1-z})
\nonumber \\ && \qquad
          +\zeta(3)(30-\frac{12}{z}-12z+ 59z^2-12z^3+\frac{4}{1+z}
            - \frac{10}{1-z})
\nonumber \\ && \qquad
       + \delta(1-z)  \{
           \frac{11141}{54}
          - \frac{214}{3}\zeta(2)
          + 5\zeta(2)^2
          - \frac{854}{9}\zeta(3)
          \}
\nonumber \\ && \qquad
       - \frac{56}{3}\ln^3(1-z)  (
           2
          - \frac{1}{z}
          - z
          + z^2
          - \frac{1}{1-z}
          )
\nonumber \\ && \qquad
       -2 \ln^2(1-z)  (
          16
          +\frac{19}{3z}
          +z
          - \frac{19}{3}z^2
          -\frac{11}{1-z}
          )
\nonumber \\ && \qquad
       - \ln(1 - z){\rm Li}_2(1 - z)  (
           102
          + \frac{12}{z}
          + 76z
          + 33z^2
          - 4z^3
          - \frac{16}{1+z}
\nonumber \\ && \qquad
          - \frac{2}{1-z}
          )
       -\frac{1}{9} \ln(1 - z)  (
           233
          +\frac{14}{z}
          -703z
          +148z^2
          +\frac{506}{1-z}
          )
\nonumber \\ && \qquad
         -\ln(1-z)\zeta(2)(26- \frac{12}{z}- 12z+ 25z^2- 4z^3 
             -\frac{18}{1-z})
\nonumber \\ && \qquad
       -4 \ln z\ln^2(1-z)  (
          22
          -\frac{3}{z}
          +5z
          +7z^2
          -\frac{2}{1+z}
          -\frac{5}{1-z}
          )
\nonumber \\ && \qquad
       -\frac{1}{3}\ln z\ln(1 - z)  (
          461
          +\frac{26}{z}
          +167z
          +64z^2
          -\frac{97}{1-z}
          )
\nonumber \\ && \qquad
       -4(
           7
          + \frac{4}{z}
          + 4z
          + 4z^2
          - \frac{2}{1+z}
          )\{2\ln(1+z){\rm Li}_2(-z)+\ln(1+z)\zeta(2)
\nonumber \\ && \qquad
+\ln z\ln^2(1+z)\}
       - \ln^2 z\ln(1 - z)  (
           51
          - \frac{10}{z}
          + 38z
          + \frac{17}{2}z^2
          -2z^3
\nonumber \\ && \qquad
          + \frac{4}{1+z}
          - \frac{5}{1-z}
          )
       -2 \ln^2 z\ln(1 + z)  (
          5
          +\frac{2}{z}
          +2z
          +2z^2
          -\frac{4}{1+z}
          )
\nonumber \\ && \qquad
       - 2\ln^3 z  (
          \frac{8}{3z}
          +12z
          - \frac{14}{3}z^2
          +\frac{1}{1+z}
          + \frac{11}{3}\frac{1}{1-z}
          )
       -2 \ln^2 z  (
          44
          +\frac{7}{z}
\nonumber \\ && \qquad
          +\frac{97}{3}z
          + \frac{11}{3}z^2
          - \frac{31}{15}z^3
          -\frac{11}{1-z}
          )
       -2 \ln z{\rm Li}_2(1 - z)  (
          54 -\frac{20}{z}
          +12z
\nonumber \\ && \qquad
          +25z^2
          - 4z^3
          +\frac{4}{1+z}
          -\frac{14}{1-z}
          )
       - 4\ln z{\rm Li}_2( - z)  (
           3
          + \frac{2}{z}
          + 2z
          + 2z^2
\nonumber \\ && \qquad
          - \frac{2}{1+z}
          + \frac{2}{1-z}
          )
       - \ln z  (
           \frac{1727}{15}
          + \frac{3598}{45z}
          + \frac{564}{5}z
          + \frac{848}{45}z^2
          + \frac{521}{9}\frac{1}{1-z}
          )
\nonumber \\ && \qquad
        +\ln z\zeta(2)(6 - \frac{12}{z}- 28z+ 25z^2- 4z^3
          - \frac{10}{1-z})
       -2{\rm Li}_2(1 - z)  (
          67
\nonumber \\ && \qquad
- \frac{28}{3z}
          +13z
          + \frac{82}{3}z^2
          + \frac{13}{3}\frac{1}{1-z}
          )
       +\frac{2}{3}(
           17
          - \frac{32}{5}\frac{1}{z^2}
          + \frac{10}{z}
          + 11z
          - 2z^2
\nonumber \\ && \qquad
          - \frac{62}{5}z^3
          )\{\ln z\ln(1+z)+{\rm Li}_2( - z)\}
       + {\rm Li}_3(1 - z)  (
           114
          + \frac{4}{z}
          + 68z+67z^2
\nonumber \\ && \qquad
          - 12z^3
          - \frac{16}{1+z}
          - \frac{6}{1-z}
          )
       + 4{\rm Li}_3( - z)  (
           1
          + \frac{2}{z}
          + 2z
          + 2z^2
          + \frac{4}{1-z}
          )
\nonumber \\ && \qquad
       + {\rm S}_{12}(1 - z)  (
          - 126
          + \frac{68}{z}
          + 36z
          - 91z^2
          + 12z^3
          - \frac{8}{1+z}
          + \frac{46}{1-z}
          )
\nonumber \\ && \qquad
       -8{\rm S}_{12}( - z)  (
          11
          +\frac{6}{z}
          +6z
          +6z^2
          -\frac{4}{1+z}
          )
       + 8(2+z+z^2- \frac{1}{1+z}
\nonumber \\ && \qquad
+\frac{1}{z})
          \{-2\ln(1 + z){\rm Li}_2(1 - z)
           -2\ln z\ln(1 - z)\ln(1 + z)+{\rm S}_{12}(z^2)\}]
\nonumber \\ && \qquad
+n_fC_AT_f[
 \frac{4}{3}(43-28z)
          - \frac{356}{9}(\frac{1}{z}-z^2)
          - \frac{64}{3}\frac{1}{1-z}
       +\frac{4}{3}\zeta(2)(7+\frac{2}{z}
\nonumber \\ && \qquad
        + 16z+2z^2-\frac{6}{1-z})
       + \frac{4}{9}\delta(1-z)\{
          - \frac{806}{3}
          + 59\zeta(2)
          + 34\zeta(3)
          \}
\nonumber \\ && \qquad
       +8 \ln^2(1-z)  (
          2
          - \frac{1}{z}
          -z
          + z^2
          - \frac{1}{1-z}
          )
       - \frac{4}{9}\ln(1 - z)(
           41
          - \frac{30}{z}
\nonumber \\ && \qquad
          -7z
          + 30z^2
          - \frac{34}{1-z}
          )
       + \frac{4}{3}\ln z\ln(1 - z)  (
           25
          - \frac{12}{z}
          - 14z
          + 12z^2
\nonumber \\ && \qquad
          - \frac{8}{1-z}
          )
       + 4\ln^2 z  (
           5
          - \frac{8}{3z}
          - 3z
          + 2z^2
          - \frac{2}{1-z}
          )
\nonumber \\ && \qquad
       -4\ln z  (
          5
          -\frac{4}{9z}
          - 2z
          + \frac{10}{3}z^2
          - \frac{37}{9}\frac{1}{1-z}
          )
       + \frac{8}{3}{\rm Li}_2(1 - z)  (
           3
          - \frac{2}{z}
          - 6z
\nonumber \\ && \qquad
          + 2z^2
          + \frac{2}{1-z}
          )
       + \frac{16}{3}(
           3
          + \frac{1}{z}
          + 3z
          +z^2
          )\{\ln z\ln(1+z)+{\rm Li}_2( - z)\}
\nonumber \\ && \qquad
       + 8z^2\{\ln(1 - z){\rm Li}_2(1 - z)+\ln(1-z)\zeta(2)
             +\frac{1}{2}\ln^2 z\ln(1 - z)
\nonumber \\ && \qquad
+2\ln z{\rm Li}_2(1 - z)
-\ln z\zeta(2)-3\zeta(3)-3{\rm Li}_3(1 - z)+3{\rm S}_{12}(1 - z)\}
]
\nonumber \\ && \qquad
+n_fC_FT_f[
- \frac{80}{9}(1-z)
          + \frac{1720}{27}(\frac{1}{z}-z^2)
       +\delta(1-z)\{-\frac{55}{3}+16\zeta(3)\}
\nonumber \\ && \qquad
       +4\zeta(2)(1 - \frac{4}{3z}-9z- \frac{4}{3}z^2 )
       + 8\ln^2(1-z)  (
           1
          + \frac{4}{3z}
          - z
          - \frac{4}{3}z^2
          )
\nonumber \\ && \qquad
       + \ln(1 - z) \{
           \frac{160}{3}(1-z)
          - \frac{176}{9}(\frac{1}{z}-z^2)
          \}
       + 8\ln z\ln(1 - z)  (
           5
          + 3z
\nonumber \\ && \qquad
          - \frac{8}{3}z^2
          )
       + 2\ln^2 z  (
           13
          + 19z
          - \frac{16}{3}z^2
          )
       + 8\ln z  (
           \frac{35}{3}
          + 5z
          + \frac{22}{9}z^2
          )
\nonumber \\ && \qquad
       + \frac{32}{3}{\rm Li}_2(1 - z)  (
           3
          -\frac{1}{z}
          + 3z
          -z^2
          )
       - \frac{32}{3}(
           3
          +\frac{1}{z}
          + 3z
          +z^2
          )\{\ln z\ln(1+z)
\nonumber \\ && \qquad
+{\rm Li}_2( - z)\}
       + 8(1+z)\{4\ln(1 - z){\rm Li}_2(1 - z)+2\ln z\ln^2(1-z)
\nonumber \\ && \qquad
              +3\ln^2 z\ln(1 - z)
+\frac{7}{6}\ln^3 z
              +4\ln z{\rm Li}_2(1 - z)+\ln z\zeta(2)
\nonumber \\ && \qquad
              -4{\rm Li}_3(1 - z)+2{\rm S}_{12}(1 - z)\}
]
+n_f^2T_f^2\delta(1-z)[16-\frac{16}{9}\zeta(2)]
\Big]\,.
\end{eqnarray}
The unphysical part is given by
(see Eq. (\ref{eqn:2.35}))
%(A13)
\begin{eqnarray}
\label{eqn:A13}
&&\hat A_{gg}^{\rm EOM}
\Big(z, \frac{-p^2}{\mu^2},\frac{1}{\varepsilon}\Big)
= 
F \,\, \Big[C_A[-2+8z-8z^2+\frac{1}{z}(1-\hat{\xi})]
\nonumber \\ && \qquad
+\varepsilon C_A[-1+4z-4z^2+\frac{1}{2z}(1-\hat{\xi})][1+\ln(1-z)+\ln z]
\Big]
\nonumber \\ && \qquad
+ F^2\,\, \Big[
\frac{1}{\varepsilon}\Big\{
C_A^2[
 - \frac{184}{3}
          + \frac{6}{z}
          - \frac{80}{3}z
          + 56z^2
   - 8\ln z  (
           1
          + 16z
          )
\nonumber \\ && \qquad
- 16\ln(1 - z)  (
           1
          - 4z
          + 4z^2
          )
]
+n_fC_AT_f\frac{8}{3}[
 4
          - \frac{1}{z}
          - 16z
          + 16z^2
]
\nonumber \\ && \qquad
+n_fC_FT_f\frac{64}{3}[
 3
          - \frac{1}{z}
          - 2z^2+3z\ln z]\Big\}
\nonumber \\ && \qquad
+C_A^2[\frac{2}{45}(
-857
          +\frac{8}{z}
          - 28z
          + 937z^2)
       +\zeta(2)(10+\frac{8}{3z}- 25z+ 8z^2
\nonumber \\ && \qquad
- \frac{248}{15}z^3 )
       +\zeta(3)(13- 41z+ 91z^2-24z^3)
       - 4(
           1
          - 4z
          + 4z^2
          )
\nonumber \\ && \qquad
\times\{3\ln^2(1-z)+\ln z{\rm Li}_2(-z)\}
       - \ln(1 - z)  (
           \frac{241}{3}
          - \frac{10}{3z}
          - 27z
          - \frac{40}{3}z^2
          )
\nonumber \\ && \qquad
+\ln(1-z)\zeta(2)(1+3z-9z^2+8z^3)
       - \ln z\ln(1 - z)  (
           34
          + \frac{8}{3z}
          + 43z
\nonumber \\ && \qquad
          + 72z^2
          )
       + (
           \frac{29}{3}
          + \frac{9}{5}\frac{1}{z^2}
          - 4z
          - 32z^2
          - \frac{248}{15}z^3
          )\{\ln z\ln(1 + z)+{\rm Li}_2(-z)\}
\nonumber \\ && \qquad
       -2 \ln^2 z  (
          3
          +47z
          - \frac{62}{15}z^3
          )
-2 {\rm Li}_2(1 - z)  (
          22
          +\frac{8}{3z}
          + 11z
          + 40z^2
          )
\nonumber \\ && \qquad
       -\frac{1}{15} \ln z  (
           829
          - \frac{8}{z}
          + 2589z
          - 328z^2
          )
       +(
           3
          - 19z
          + 25z^2
          - 8z^3
          )
\nonumber \\ && \qquad
\times\{-\ln(1-z){\rm Li}_2(1-z)-\frac{1}{2}\ln^2 z\ln(1 - z)
-2\ln z{\rm Li}_2(1 -z)
\nonumber \\ && \qquad
+\ln z\zeta(2)+ 3{\rm Li}_3(1 - z)\}
    - {\rm S}_{12}(1 - z)  (
           13
          - 73z
          + 91z^2
          - 24z^3
          )
\nonumber \\ && \qquad
      + 4{\rm Li}_3( - z)  (
           3
          - 4z
          + 12z^2
          )
         +4(1+4z+4z^2)\{-2\ln(1 + z){\rm Li}_2( - z)
\nonumber \\ && \qquad
-\ln(1 + z)\zeta(2)-\ln z\ln^2(1+z)+\frac{1}{2}\ln^2 z\ln(1 + z)
-2{\rm S}_{12}( - z)\}
]
\nonumber \\ && \qquad
+n_fC_AT_f[
 \frac{88}{9}
          - \frac{28}{3z}
          + \frac{128}{9}z
          -16z^2
       + 8\ln(1 - z)  (
           \frac{7}{3}
          + \frac{1}{3z}
          - 6z
          + 4z^2
          )
\nonumber \\ && \qquad
       + \frac{4}{3}\ln z  (
           4
          -\frac{5}{z}
          -4z
          +24z^2
          )
       + 8(
           1
          + \frac{2}{3z}
          - 2z
          )\{-\zeta(2)
\nonumber \\ && \qquad
+\ln z\ln(1-z)+2{\rm Li}_2(1 - z)\}
       +16z(1-z)\{-\ln(1 - z){\rm Li}_2(1 - z)
\nonumber \\ && \qquad
-\ln(1-z)\zeta(2)
        -\frac{1}{2}\ln^2 z\ln(1 - z)
        -2\ln z{\rm Li}_2(1 - z)+\ln z\zeta(2)
\nonumber \\ && \qquad
        +3\zeta(3)+3{\rm Li}_3(1 - z)-3{\rm S}_{12}(1 - z)\}
]
\nonumber \\ && \qquad
+n_fC_FT_f[
- \frac{64}{3}
          + \frac{128}{9z}
          - \frac{32}{3}z
          + \frac{160}{9}z^2
 + 64\ln(1 - z)  (
           1
          - \frac{1}{3z}
          - \frac{2}{3}z^2
          )
\nonumber \\ && \qquad
       + 32\ln z  (
           1
          + 2z
          - \frac{4}{3}z^2
          )
  + 64z\{\frac{3}{4}\ln^2 z+\ln z\ln(1 - z)+{\rm Li}_2(1 - z)\}
]
\Big]\,.
\nonumber \\ && \qquad
\end{eqnarray}
%%%%%%%%%%%%%%%%%%%%%%%%%%%%%%%%%%%%
The part which is due to the contribution of the NGI operators $O_A$ and 
$O_{\omega}$ (see Eq. (\ref{eqn:2.36})) is equal to
%(A14)
\begin{eqnarray}
\label{A14}
&&\hat A_{gg}^{\rm NGI} 
\Big(z, \frac{-p^2}{\mu^2},\frac{1}{\varepsilon}\Big)
= 
F \,\, C_A\Big[
\frac{1}{\varepsilon}[4-\frac{4}{z}]
+2(1-\frac{1}{z})[1+\ln z+\ln(1-z)]
\nonumber \\ && \qquad
+\varepsilon (1-\frac{1}{z})[\frac{1}{2}\zeta(2)+\ln(1-z)+\ln z
+\frac{1}{2}\{\ln z+\ln(1-z)\}^2]\Big]
\nonumber \\ && \qquad
+ F^2 \,\, \Big[
\frac{1}{\varepsilon^2}\Big\{C_A^2[
\frac{4}{3}(1+2z^2)
          + \frac{4}{z}
          - 8z
  + 22\ln(1 - z)  (
           1
          - \frac{1}{z}
          )
\nonumber \\ && \qquad
       + 2\ln z  (
           5
          + \frac{6}{z}
          )
]
-n_fC_AT_f\frac{32}{3}(1-\frac{1}{z})\Big\}
\nonumber \\ && \qquad
+\frac{1}{\varepsilon}\Big\{
C_A^2[
 - \frac{73}{9}
          + \frac{250}{9z}
          - \frac{67}{3}z
          + \frac{8}{3}z^2
 + \frac{33}{2}\ln^2(1-z)  (
           1
          -\frac{1}{z}
          )
\nonumber \\ && \qquad
       + \frac{8}{3}\ln(1 - z)  (
          1
          +\frac{1}{z}
          -3z
          +z^2
          )
       + 2\ln z\ln(1 - z)  (
           15
          - \frac{4}{z}
          )
\nonumber \\ && \qquad
      + \frac{1}{2}\ln^2 z  (
          15
          + \frac{14}{z}
          )
       + \frac{2}{3}\ln z  (
           65
          + \frac{18}{z}
          + 6z
          )
       + 2{\rm Li}_2(1 - z)  (
           14
          - \frac{3}{z}
          )
\nonumber \\ && \qquad
       + 4(1+\frac{1}{z})\{\frac{1}{2}\zeta(2)
+\ln z\ln(1 + z)+{\rm Li}_2( - z)\}
]
\nonumber \\ && \qquad
-n_fC_AT_f\frac{16}{3}(1-\frac{1}{z})[\frac{1}{3}+2\ln z+\ln(1-z)]\Big\}
\nonumber \\ && \qquad
+C_A^2[
 - \frac{59}{27}
          + \frac{1837}{54z}
          - \frac{61}{2}z
          - \frac{4}{3}z^2
     + \zeta(2)(\frac{1}{3}+ \frac{2}{z}-5z+ \frac{2}{3}z^2)
\nonumber \\ && \qquad
          +\zeta(3)(5+\frac{2}{z}-18z+ 9z^2)
       + \frac{77}{12}\ln^3(1-z)  (
           1
          - \frac{1}{z}
          )
\nonumber \\ && \qquad
       + \ln^2(1-z)  (
          \frac{5}{3}
          +\frac{1}{z}
          - 4z
          + \frac{4}{3}z^2
          )
       + \ln(1 - z){\rm Li}_2(1 - z)  (
           29
          - \frac{2}{z}
          + 6z
\nonumber \\ && \qquad
          - 3z^2
          )
       + \frac{1}{9}\ln(1 - z)  (
           1
          +\frac{149}{z}
          -174z
          +24z^2
          )
\nonumber \\ && \qquad
        -\frac{1}{2}\ln(1-z)\zeta(2)(1+\frac{5}{z}-12z+6z^2)
       + \ln z\ln^2(1-z)  (
           22
          - \frac{7}{z}
          )
\nonumber \\ && \qquad
       + \frac{1}{3}\ln z\ln(1 - z)  (
           134
          +\frac{23}{z}
          +21z
          )
       + \frac{1}{2}\ln^2 z\ln(1 - z)  (
           33
          - \frac{7}{z}
          + 6z
\nonumber \\ && \qquad
          - 3z^2
          )
       + \ln^3 z  (
           \frac{35}{12}
          + \frac{5}{2z}
          )
       + \ln^2 z  (
           \frac{193}{6}
          +\frac{9}{z}
          + 5z
          - \frac{2}{3}z^2
          )
\nonumber \\ && \qquad
       + 2 \ln z {\rm Li}_2(1 - z)  (
           13
          - \frac{7}{z}
          + 6z
          - 3z^2
          )
       + \ln z  (
           \frac{985}{18}
          + \frac{412}{9z}
          + \frac{71}{6}z
          )
\nonumber \\ && \qquad
 +\ln z\zeta(2)(\frac{9}{2}+ \frac{4}{z}- 6z+ 3z^2 )
       + 2{\rm Li}_2(1 - z)  (
           22
          + \frac{4}{3z}
          + 7z
          - \frac{2}{3}z^2
          )
\nonumber \\ && \qquad
       - (
          3
          - \frac{1}{z^2}
          + \frac{2}{z}
          )\{\ln z\ln(1+z)+{\rm Li}_2( - z)\}
       - {\rm Li}_3(1 - z)  (
           23
          - \frac{2}{z}
          + 18z
\nonumber \\ && \qquad
          - 9z^2
          )
       + {\rm S}_{12}(1 - z)  (
           15
          - \frac{17}{z}
          + 18z
          - 9z^2
          )
          -2(1+\frac{1}{z})
            \{-\ln z{\rm Li}_2( - z)
\nonumber \\ && \qquad
-2\ln(1 + z)\zeta(2)
-4\ln(1+z){\rm Li}_2(-z)
-2\ln(1 + z){\rm Li}_2(1-z)
\nonumber \\ && \qquad
-2\ln z\ln(1 - z)\ln(1 + z)
              -2\ln z\ln^2(1+z)-\frac{1}{2}\ln^2 z\ln(1 + z)
\nonumber \\ && \qquad
+{\rm Li}_3( - z)
              -6{\rm S}_{12}( - z)
+{\rm S}_{12}(z^2)\}]
\nonumber \\ && \qquad
-n_fC_AT_f\frac{4}{3}(1-\frac{1}{z})[
        \frac{32}{9}
       +\frac{2}{3}\{\ln(1-z)+2\ln z\}+2\zeta(2)
\nonumber \\ && \qquad
       +\{2\ln z+\ln(1-z)\}^2]\Big]\,.
\end{eqnarray}
%\end{appendixOB A}
%%%% end of Appendix A%%%%
%\newcommand{\mysection}{\setcounter{equation}{0}\section}
%\newcommand{\ps}{p \hspace{-0.52em}/\hspace{0.1em}}
%\newcommand{\Ds}{\Delta \hspace{-0.52em}/\hspace{0.1em}}
%\mysection*{Appendix B}
%\begin{appendix B}
%\appendix{Appendix B}
%\setcounter{section}{2}
\section{}

In this Appendix we present the non-gauge invariant (NGI) operators with
their corresponding operator matrix elements (OME's). They are needed for the
renormalization of the physical operators due to the mixing between
them discussed in section 2. Two of them are given in Eqs. (2.9) and
(2.10) of \cite{hane} (for their construction see \cite{dita}). These 
operators, which already show up in the case where only the gluonic operator
in Eq. (\ref{eqn:2.3}) is present, are given by
\begin{eqnarray}
\label{eqn:B1}
&& O_A^{\mu_1,\mu_2 \cdots \mu_n}(x) = i^{n-2} {\cal S} \Big [ 
F_{a,\alpha}^{\mu_1}(x) D^{\alpha} \partial^{\mu_2} \cdots \partial^{\mu_{n-1}}
A_a^{\mu_n}(x)
\nonumber\\[2ex]
&& + i g f_{abc} F_{a,\alpha}^{\mu_1}(x) \sum_{i=2}^{n-1} \kappa_i
\partial^{\alpha} \Big \{ \Big (\partial^{\mu_2} \cdots \partial^{\mu_{i-2}}
A_b^{\mu_{i-1}} (x) \Big ) \Big ( \partial^{\mu_{i}} \cdots 
\partial^{\mu_{n-1-i}}A_c^{\mu_n} (x) \Big ) \Big \}
\nonumber\\[2ex]
&&  + {\cal O}(g^2) \Big ] \,,
\end{eqnarray}
and
\begin{eqnarray}
\label{eqn:B2}
&& O_{\omega}^{\mu_1,\mu_2 \cdots \mu_n} (x) = i^{n-2} {\cal S} \Big [
\xi_a(x) \partial^{\mu_1} \cdots \partial^{\mu_n} {\bar \omega}_a
\nonumber\\[2ex]
&& - i g f_{abc} \xi_a(x) \sum_{i=2}^{n-1} \eta_i
\partial^{\mu_1} \Big \{ \Big (\partial^{\mu_2} \cdots \partial^{\mu_{i-1}}
{\bar \omega}_b (x) \Big ) \Big ( \partial^{\mu_{i}} \cdots
\partial^{\mu_{n-1-i}} A_c^{\mu_n} (x) \Big ) \Big \} 
\nonumber\\[2ex]
&& + {\cal O}(g^2) \Big ] \,,
\end{eqnarray}
where $\xi$ and $\bar \omega$ are the ghost and antighost respectively.
In the expressions above $\eta$ is defined in Eq. (\ref{eqn:2.45}). Further
$\kappa_i$ and $\eta_i$ are calculated in \cite{hane} and are given by
\begin{eqnarray}
\label{eqn:B3}
\kappa_i=  \frac{1}{8} (-1)^i + \frac{3}{8}
\frac{(n-2)!}{(i-1)!(n-i-1)!}- \frac{3}{8}\frac{(n-2)!}{i!(n-i-2)!} \,,    
\end{eqnarray}
and 
\begin{eqnarray}
\label{eqn:B4}
\eta_i=  \frac{1}{4} (-1)^i + \frac{3}{4}
\frac{(n-2)!}{(i-1)!(n-i-1)!}+\frac{1}{4}\frac{(n-2)!}{i!(n-i-2)!} \,,
\end{eqnarray}
respectively.
If we also include the quark singlet operators in Eq. (\ref{eqn:2.2})
then we have to add
\begin{eqnarray}
\label{eqn:B5}
O_B^{\mu_1,\mu_2 \cdots \mu_n}(x) = i^{n-1} {\cal S} \Big [g {\bar \psi}_k(x)
\gamma^{\mu_1} (T_a)_{kl} A_a^{\mu_2} (x) \partial^{\mu_3} \cdots 
\partial^{\mu_n} \psi_l(x) + {\cal O}(g^3) \Big ] \,.
\end{eqnarray}
Notice that the above operators are not BRS-exact in the strict sense of
\cite{brs} (see also \cite{cosc}). This might affect the non-logarithmic
terms in the renormalized OME's which we do not need here. Further 
the operators in the above equations are corrected up to order
$g^2$. This is sufficient to get finite two-loop OME's. However in order
to carry out the renormalization on the three-loop level one has to compute 
higher order corrections to  Eqs. (\ref{eqn:B1}), (\ref{eqn:B2})
and (\ref{eqn:B5}). The 
operator vertices corresponding to $O_A$ in Eq. (\ref{eqn:B1}) and $O_B$ 
in Eq. (\ref{eqn:B2}), albeit sandwiched between gluon states, are 
presented in Appendix A.3 of \cite{hane}. In section 2 we need the 
$O_B$ operator vertex when it is sandwiched between quark states. In this 
case we have to compute the quark-quark-gluon vertex which is given by
\begin{eqnarray}
\label{eqn:B6}
V_{a,kl}^{\mu} = g (T_a)_{kl}
{\Delta \hspace{-0.52em}/\hspace{0.1em}}
%\Ds 
\Delta^{\mu} (\Delta \cdot k)^{n-2}\,,
\end{eqnarray}
where $k$ stands for the momentum of the gluon.

We now list the OME's needed for the renormalization of the physical operators
in section 2 (for the notation see Appendix A).
The two OME's referring to the NGI operators $O_A$ and $O_B$, when sandwiched
between quarks states, can be split according to Eq. (\ref{eqn:2.7}).
The expressions in (\ref{eqn:2.46}) and (\ref{eqn:2.47}) become 
\begin{eqnarray}
\label{eqn:B7}
&&\hat A_{Aq}^{\rm PHYS}
\Big(z, \frac{-p^2}{\mu^2},\frac{1}{\varepsilon}\Big)
=\,F\,C_F\Big[
\frac{1}{\varepsilon}\Big\{-8+8\frac{1}{z}\Big\}
-2+2\frac{1}{z}+(-4+4\frac{1}{z})[\ln(1-z)
\nonumber \\ && \qquad
+\ln z] -(1-\hat{\xi})[2-\frac{3}{2z}]-\varepsilon\Big\{(1-\frac{1}{z})[
\ln(1-z)+\ln z+\{\ln(1-z)+\ln z\}^2
\nonumber \\ && \qquad
+\zeta(2)]
+(1-\hat{\xi})[\frac{1}{2}-\frac{1}{4z}+(1-
\frac{3}{4z})\{\ln(1-z)+\ln z\}]
\Big\}\Big]\,,
\end{eqnarray}
and
\begin{eqnarray}
\label{eqn:B8}
&&\hat A_{Aq}^{\rm EOM}
\Big(z, \frac{-p^2}{\mu^2},\frac{1}{\varepsilon}\Big)
=(1-\hat{\xi})\,F\,C_F\Big[1
+\frac{\varepsilon}{2}[1+\ln(1-z)+\ln z]\Big]\,,
\end{eqnarray}
respectively.
Likewise we obtain for (\ref{eqn:2.48}) and (\ref{eqn:2.49})
\begin{eqnarray}
\label{eqn:B9}
&&\hat A_{Bq}^{\rm PHYS}
\Big(z, \frac{-p^2}{\mu^2},\frac{1}{\varepsilon}\Big)
= \, F \,C_F\Big[
\frac{1}{\varepsilon}\Big\{8-8\frac{1}{z}\Big\}
+(4-4\frac{1}{z})[\ln(1-z)+\ln z]
\nonumber \\ && \qquad
+(1-\hat{\xi})
(2-\frac{2}{z})
+\varepsilon(1-\frac{1}{z})\Big\{
\{\ln(1-z)+\ln z\}^2
+\zeta(2)
+(1-\hat{\xi})
\nonumber \\ && \qquad
\times[\ln(1-z)+\ln z]
\Big\}\Big] \,,
\end{eqnarray}
and
\begin{eqnarray}
\label{eqn:B10}
&&\hat A_{Bq}^{\rm EOM}
\Big(z, \frac{-p^2}{\mu^2},\frac{1}{\varepsilon}\Big)
=0\,,
\end{eqnarray}
respectively. 
Next we sandwich the NGI operators $O_A$ and $O_{\omega}$ between gluon
states. Decomposing the OME's according to (\ref{eqn:2.53}) we get for
$\hat A_{Ag}$ the following results. The physical and unphysical parts (see
Eq.  (\ref{eqn:2.54}) and (\ref{eqn:2.55})) become equal to 
\begin{eqnarray}
\label{eqn:B11}
&&\hat A_{Ag}^{\rm PHYS}
\Big(z, \frac{-p^2}{\mu^2},\frac{1}{\varepsilon}\Big)
= \,F\,C_A\Big[
\frac{1}{\varepsilon}\Big\{2z-2z^2+(1-\hat{\xi})(12-\frac{2}{z}-3z-2z^2)
\nonumber \\ && \qquad
+(1-\hat{\xi})^2(\frac{1}{2}
-z-\frac{z^2}{2})\Big\}
+z(1-z)[\ln(1-z)+\ln z]+2-4\frac{1}{z}
\nonumber \\ && \qquad
+(1-\hat{\xi})[\frac{7}{2}+\frac{1}{z^2}
-\frac{5}{2z}-\frac{z}{2}
+(6-\frac{1}{z}-\frac{3}{2}z-z^2)\{\ln(1-z)+\ln z\}]
\nonumber \\ && \qquad
+(1-\hat{\xi})^2[\frac{7}{8}-\frac{3}{4}z
+\frac{1}{4}(1-2z-z^2)\{\ln(1-z)+\ln z\}]
\nonumber \\ && \qquad
+\varepsilon\Big\{(1-\frac{2}{z})[\ln(1-z)+\ln z]+\frac{1}{4}z(1-z)
[\{\ln(1-z)+\ln z\}^2
+\zeta(2)]
\nonumber \\ && \qquad
+(1-\hat{\xi})[-\frac{1}{2z}+(\frac{7}{4}+\frac{1}{2z^2}-\frac{5}{4z}
-\frac{1}{4}z)\{\ln(1-z)+\ln z\}
\nonumber \\ && \qquad
+(\frac{3}{2}-\frac{1}{4z}-\frac{3z}{8}-\frac{z^2}{4})\{(\ln(1-z)+\ln z)^2
+\zeta(2)\}]
+(1-\hat{\xi})^2[\frac{1}{16}
\nonumber \\ && \qquad
+(\frac{7}{16}-\frac{3}{8}z)\{\ln(1-z)+\ln z\}
+\frac{1}{16}(1-2z-z^2)\{(\ln(1-z)+\ln z)^2
\nonumber \\ && \qquad
+\zeta(2)\}]
\Big\}\Big]\,,
\end{eqnarray}
and
\begin{eqnarray}
\label{eqn:B12}
&&\hat A_{Ag}^{\rm EOM}
\Big(z, \frac{-p^2}{\mu^2},\frac{1}{\varepsilon}\Big)
= \,F\,C_A\Big[-\frac{1}{\varepsilon}13\delta(1-z)(1-\hat{\xi})
+4+2z-2z^2+(1-\hat{\xi})
\nonumber \\ && \qquad
\times[5-\frac{1}{z^2}+\frac{5}{2z}-2z-2z^2+\delta(1-z)
-\frac{1}{2}\frac{1}{1-z}-\frac{3}{2}\frac{1}{(1-z)^2}]
\nonumber \\ && \qquad
+(1-\hat{\xi})^2[\frac{3}{2}-\frac{1}{2z}-2z-\frac{z^2}{2}
-\frac{27}{8}\delta(1-z)-\frac{3}{4}\frac{1}{1-z}]
\nonumber \\ && \qquad
+\varepsilon\Big\{(2+z-z^2)[\ln(1-z)+\ln z]+(1-\hat{\xi})[\frac{5}{2}
-\frac{z}{2}-\frac{1}{1-z}
\nonumber \\ && \qquad
+\delta(1-z)\{-\frac{15}{2}+\frac{11}{8}\zeta(2)\}
+(\frac{5}{2}-\frac{1}{2z^2}+\frac{5}{4z}-z-z^2
\nonumber \\ && \qquad
-\frac{1}{4}\frac{1}{1-z}-\frac{3}{4}\frac{1}{(1-z)^2})
\{\ln(1-z)+\ln z\}]
\nonumber \\ && \qquad
+(1-\hat{\xi})^2[\frac{3}{4}-\frac{1}{8z^2}+\frac{3}{16z}-\frac{3z}{4}
+\frac{3}{2}\delta(1-z)\{1-\frac{1}{4}\zeta(2)\}
\nonumber \\ && \qquad
-\frac{3}{16}\frac{1}{(1-z)^2}
-\frac{3}{8}\frac{1}{1-z}
+(\frac{3}{4}-\frac{1}{4z}-z-\frac{z^2}{4}-\frac{3}{8}\frac{1}{1-z})
\nonumber \\ && \qquad
\times\{\ln(1-z)+\ln z\}
]\Big\}\Big]\,,
\end{eqnarray}
respectively. The piece coming from the NGI operator $O_A$ itself and the part
due to the violation of the Ward-identity (WI) (see Eqs. (\ref{eqn:2.56}),
 (\ref{eqn:2.57})) are given by
\begin{eqnarray}
\label{eqn:B13}
&&\hat A_{Ag}^{\rm NGI}
\Big(z, \frac{-p^2}{\mu^2},\frac{1}{\varepsilon}\Big)
=\,F\,C_A\Big[
\frac{1}{\varepsilon}\Big\{2-5z+2z^2+\frac{4}{z}-6\frac{1}{1-z}
-\frac{3}{2}\delta(1-z)
\nonumber \\ && \qquad
+(1-\hat{\xi})[\frac{61}{2}-\frac{21}{2z}
-\frac{21z}{2}-z^2+\frac{145}{24}\delta(1-z)
+\frac{5}{4}\frac{1}{1-z}]
\nonumber \\ && \qquad
+(1-\hat{\xi})^2[-\frac{3}{2}-\frac{1}{2z}
+\frac{11z}{4}-\frac{5}{2}z^2+\frac{11}{12}\delta(1-z)
-\frac{3}{4}\frac{1}{1-z}]
\Big\}
\nonumber \\ && \qquad
+(1-\frac{5}{2}z+z^2+\frac{2}{z}-\frac{3}{1-z})[\ln(1-z)+\ln z]
\nonumber \\ && \qquad
-\frac{5}{2}+\frac{2}{z}
+\delta(1-z)[\frac{4}{3}-3\zeta(2)]+(1-\hat{\xi})[\frac{17}{2}+\frac{1}{8z^2}
-\frac{41}{16z}-3z
\nonumber \\ && \qquad
+\delta(1-z)\{-\frac{1}{72}+\frac{5}{8}\zeta(2)\}
+\frac{7}{16}\frac{1}{1-z}+\frac{3}{16}\frac{1}{(1-z)^2}
+(\frac{61}{4}
\nonumber \\ && \qquad
-\frac{21}{4z}-\frac{21z}{4}-\frac{z^2}{2}+\frac{5}{8}\frac{1}{1-z})
\{\ln(1-z)+\ln z\}
]
\nonumber \\ && \qquad
+(1-\hat{\xi})^2[\frac{7}{4}+\frac{1}{8z^2}
-\frac{15}{16z}-\frac{1}{2}z
+\delta(1-z)\{\frac{97}{36}-\frac{3}{8}\zeta(2)\}
\nonumber \\ && \qquad
+\frac{11}{16}\frac{1}{1-z}+\frac{3}{16}\frac{1}{(1-z)^2}
+(-\frac{3}{4}
-\frac{1}{4z}+\frac{11}{8}z-\frac{5}{4}z^2-\frac{3}{8}\frac{1}{1-z})
\nonumber \\ && \qquad
\times\{\ln(1-z)
+\ln z\}
]
+\varepsilon\Big\{\delta(1-z)[-\frac{23}{18}+\frac{3}{16}\zeta(2)
\nonumber \\ && \qquad
+3\zeta(3)]
+(-\frac{5}{4}+\frac{1}{z})[\ln(1-z)+\ln z]
\nonumber \\ && \qquad
+\frac{1}{8}(2+\frac{4}{z}-5z+2z^2-\frac{3}{4}\frac{1}{1-z})
[\{\ln(1-z)+\ln z\}^2+\zeta(2)]
\nonumber \\ && \qquad
+(1-\hat{\xi})[-\frac{1}{2}+\frac{1}{8z}
+\frac{7}{32}\frac{\ln z}{1-z}
+\frac{7}{32}\frac{\ln(1-z)}{1-z}
+\frac{1}{2}\frac{1}{1-z}
\nonumber \\ && \qquad
+\delta(1-z)[\frac{677}{432}-\frac{103}{192}\zeta(2)
-\frac{5}{8}\zeta(3)]
+(\frac{17}{4}+\frac{1}{16z^2}-\frac{41}{32z}-\frac{3}{2}z
\nonumber \\ && \qquad
+\frac{3}{32}\frac{1}{(1-z)^2})[\ln(1-z)+\ln z]
+\frac{1}{16}(61-\frac{21}{z}-21z-2z^2+\frac{5}{2}\frac{1}{1-z})
\nonumber \\ && \qquad
\times
[\{\ln z
+\ln(1-z)\}^2+\zeta(2)]
+(1-\hat{\xi})^2[\frac{5}{16}+\frac{1}{16z^2}-\frac{5}{16z}
+\frac{3}{32}\frac{1}{(1-z)^2}
\nonumber \\ && \qquad
+\frac{11}{32}\frac{\ln z}{1-z}
+\frac{11}{32}\frac{\ln(1-z)}{1-z}
+\frac{15}{32}\frac{1}{1-z}
+\delta(1-z)[-\frac{271}{432}+\frac{11}{48}\zeta(2)
+\frac{3}{8}\zeta(3)]
\nonumber \\ && \qquad
+(\frac{7}{8}+\frac{1}{16z^2}-\frac{15}{32z}-\frac{1}{4}z
+\frac{3}{32}\frac{1}{(1-z)^2})\{\ln(1-z)+\ln z\}
\nonumber \\ && \qquad
+\frac{1}{16}(-3-\frac{1}{z}+\frac{11}{2}z-5z^2-\frac{3}{2}\frac{1}{1-z})
[\{\ln(1-z)+\ln z\}^2+\zeta(2)]
]
\Big\}\Big]\,,
\nonumber \\ && \qquad
\end{eqnarray}
and
\begin{eqnarray}
\label{eqn:B14}
&&\hat A_{Ag}^{\rm WI}
\Big(z, \frac{-p^2}{\mu^2},\frac{1}{\varepsilon}\Big)
=\,F\, C_A\Big[
\frac{1}{\varepsilon}\Big\{(1-\hat{\xi})[5-z-\frac{2}{z}
-\frac{5}{8}\delta(1-z)]
+(1-\hat{\xi})^2[-4
\nonumber \\ && \qquad
+\frac{5}{2z}+z
+\frac{1}{4}\delta(1-z)
+\frac{1}{4}\frac{1}{1-z}]
\Big\}
+\frac{1}{2}
+(1-\hat{\xi})[\frac{1}{4}-\frac{1}{8z^2}
-\frac{11}{16z}
\nonumber \\ && \qquad
+\frac{3}{8}\delta(1-z)
+\frac{7}{16}\frac{1}{1-z}-\frac{3}{16}\frac{1}{(1-z)^2}
+(\frac{5}{2}
-\frac{1}{z}-\frac{z}{2})\{\ln(1-z)
\nonumber \\ && \qquad
+\ln z\}]
+(1-\hat{\xi})^2[-\frac{5}{8}+\frac{1}{z}
+\delta(1-z)\{-\frac{9}{16}+\frac{1}{8}\zeta(2)\}
-\frac{1}{16}\frac{1}{1-z}
\nonumber \\ && \qquad
+(-2
+\frac{5}{4z}+\frac{z}{2}+\frac{1}{8}\frac{1}{1-z})\{\ln(1-z)+\ln z\}
]
\nonumber \\ && \qquad
+\varepsilon\Big\{\frac{1}{4}[\ln(1-z)+\ln z]
+(1-\hat{\xi})[-\frac{1}{8z}
+\frac{7}{32}\frac{\ln z}{1-z}
+\frac{7}{32}\frac{\ln(1-z)}{1-z}
\nonumber \\ && \qquad
+\delta(1-z)[-\frac{15}{16}+\frac{19}{64}\zeta(2)]
+(\frac{1}{8}-\frac{1}{16z^2}-\frac{11}{32z}
-\frac{3}{32}\frac{1}{(1-z)^2})\{\ln z
\nonumber \\ && \qquad
+\ln(1-z)\}
+\frac{1}{8}(5-\frac{2}{z}-z)(\{\ln(1-z)+\ln z\}^2+\zeta(2))
]
\nonumber \\ && \qquad
+(1-\hat{\xi})^2[\frac{1}{32z}
-\frac{3}{32}\frac{1}{1-z}
-\frac{1}{32}\frac{\ln z}{1-z}
-\frac{1}{32}\frac{\ln(1-z)}{1-z}
\nonumber \\ && \qquad
+\delta(1-z)[\frac{3}{8}-\frac{1}{16}\zeta(2)
-\frac{1}{8}\zeta(3)]
+(-\frac{5}{16}+\frac{1}{2z})\{\ln(1-z)+\ln z\}
\nonumber \\ && \qquad
+\frac{1}{16}(-8+\frac{5}{z}+2z+\frac{1}{2}\frac{1}{1-z})
[\{\ln(1-z)+\ln z\}^2+\zeta(2)]
]
\Big\}\Big]\,,
\end{eqnarray}
respectively.\\
The last OME due to $O_{\omega}$ can be also split into four pieces
according to Eq. (\ref{eqn:2.58}). For Eq. (\ref{eqn:2.59}) we get
\begin{eqnarray}
\label{eqn:B15}
&&\hat A_{\omega g}^{\rm PHYS}
\Big(z, \frac{-p^2}{\mu^2},\frac{1}{\varepsilon}\Big)
=-\,F\,C_Az(1-z)\Big[
\frac{2}{\varepsilon}+\ln(1-z)+\ln z
\nonumber \\ && \qquad
+\varepsilon\frac{1}{4}
[\{\ln(1-z)+\ln z\}^2
+\zeta(2)]\Big]\,.
\end{eqnarray}
The expression in Eq. (\ref{eqn:2.60}) becomes
\begin{eqnarray}
\label{eqn:B16}
&&\hat A_{\omega g}^{\rm EOM}
\Big(z, \frac{-p^2}{\mu^2},\frac{1}{\varepsilon}\Big)
=-\,F\,C_Az(1-z)\Big[2
+\varepsilon[\ln(1-z)+\ln z]\Big]\,.
\end{eqnarray}
For the contribution due to the NGI operator $O_\omega$ we get (see Eq. 
(\ref{eqn:2.61})) 
\begin{eqnarray}
\label{eqn:B17}
&&\hat A_{\omega g}^{\rm NGI}
\Big(z, \frac{-p^2}{\mu^2},\frac{1}{\varepsilon}\Big)
=\,F\,C_A\Big[
\frac{1}{\varepsilon}\Big\{-2+5z-2z^2
+\frac{1}{6}\delta(1-z)\Big\}
\nonumber \\ && \qquad
+(-1+\frac{5}{2}z-z^2)[\ln(1-z)+\ln z]
+\frac{1}{2}
-\frac{2}{9}\delta(1-z)
\nonumber \\ && \qquad
+\varepsilon\Big\{\delta(1-z)[\frac{13}{54}-\frac{1}{48}\zeta(2)]
+\frac{1}{4}[\ln(1-z)+\ln z]
\nonumber \\ && \qquad
+\frac{1}{8}(-2+5z-2z^2)[\{\ln(1-z)+\ln z\}^2+\zeta(2)]
\Big\}\Big]\,.
\end{eqnarray}
Finally the contribution due to the breakdown of the Ward identity is 
(see Eq. (\ref{eqn:2.62}))
\begin{eqnarray}
\label{eqn:B18}
&&\hat A_{\omega g}^{\rm WI}
\Big(z, \frac{-p^2}{\mu^2},\frac{1}{\varepsilon}\Big)
=-\,F\,C_A\Big[\frac{1}{2}
+\varepsilon\frac{1}{4}[\ln(1-z)+\ln z]\Big]\,.
\end{eqnarray}
%\end{document}
%\end{appendix B}
%%%%%%%%%%%%%%%%%%%%% REFERENCES %%%%%%%%%%%%%%%%%%%%%%%%%%%%%%%%%%%%%%%%%%%%%
%\begin{thebibliography}{99}

\end{document}